\documentclass{aa} 

\usepackage{graphicx}
\usepackage{txfonts}
\usepackage{hyperref}
\hypersetup{
    colorlinks=true,
    linkcolor=blue,
    filecolor=magenta,      
    urlcolor=blue,
    citecolor=blue,
    pdfpagemode=FullScreen,
    }
\usepackage{graphicx}	
\usepackage{amsmath}	
\usepackage{caption}
\usepackage{subcaption}
\usepackage{multirow}
\usepackage{pifont}

\usepackage{booktabs}
\newcommand{\mass}{log(M_{\star}/M_{\odot})}
\newcommand{\massf}{log(M_{formed}/M_{\odot})}
\newcommand{\ssfr}{log(sSFR/yr^{-1})}
\newcommand{\lcdm}{\Omega_M=0.3,\: \Omega_{\Lambda}=0.7\: and\: H_0 = 70\: km\:s^{-1}\:Mpc^{-1}}
\newcommand{\cmark}{\ding{51}}
\newcommand{\xmark}{\ding{56}}
\begin{document}

   \title{A new measurement of the expansion history of the Universe at\\ z = 1.26 with cosmic chronometers in VANDELS}

   \author{E. Tomasetti\inst{1,2} \fnmsep\thanks{\email{elena.tomasetti2@unibo.it}}
          \and
          M. Moresco\inst{1,2}
          \and
          N. Borghi\inst{1,2}
          \and
          K. Jiao\inst{3,4}
          \and
          A. Cimatti\inst{1,5}
          \and
          L. Pozzetti\inst{1,2}
          \and
          A. C. Carnall\inst{6}
          \and
          \\R. J. McLure\inst{6}
          \and
          L. Pentericci\inst{7}
            }

   \institute{
            Dipartimento di Fisica e Astronomia “Augusto Righi”–Università di Bologna, via Piero Gobetti 93/2, I-40129 Bologna, Italy
            \and
            INAF - Osservatorio di Astrofisica e Scienza dello Spazio di Bologna, via Piero Gobetti 93/3, I-40129 Bologna, Italy
            \and
            Institute for Frontiers in Astronomy and Astrophysics, Beijing Normal University, Beijing 102206, People's Republic of China
            \and
            Department of Astronomy, Beijing Normal University, Beijing 100875, People's Republic of China
            \and
            INAF - Osservatorio Astrofisico di Arcetri, Largo E. Fermi 5, I-50125, Firenze, Italy
            \and
            Institute for Astronomy, University of Edinburgh, Royal Observatory, Edinburgh EH9 3HJ, UK
            \and
            INAF - Osservatorio Astronomico di Roma, Via di Frascati 33, 00078, Monte Porzio Catone, Italy
             }

   \date{}

    \titlerunning{A new measurement of the expansion history of the Universe with cosmic chronometers in VANDELS}
    \authorrunning{E. Tomasetti et al.}
 
  \abstract
   {}
   {We derive a new constraint on the expansion history of the Universe by applying the cosmic chronometers method in the VANDELS survey,
   studying the age evolution of high-redshift galaxies with a full-spectral-fitting approach.}
   {We select a sample of 39 massive ($\mathrm{\mass}$>10.8) and passive ($\mathrm{\ssfr}$<-11) galaxies from the fourth data release of the VANDELS survey at $1<z<1.5$. To minimize the potential contamination by star-forming outliers, we selected our sample by combining different selection criteria, considering both photometric and spectroscopic information. The analysis of the observed spectral features provides direct evidence of an age evolution with redshift and of mass-downsizing, with more massive galaxies presenting stronger age-related features. To estimate the physical properties of the sample, we perform full-spectral-fitting with the code \texttt{BAGPIPES}, jointly analysing spectra and photometry of our sources without any cosmological assumption on the age of the population.}
   {The derived physical properties of the selected galaxies are characteristic of a passive population, with short star formation timescales ($\mathrm{\langle \tau \rangle= 0.28 \pm 0.02 \:Gyr}$), low dust extinction ($\mathrm{\langle A_{V,dust}\rangle = 0.43\pm 0.02\: mag}$), and sub-solar metallicities ($\mathrm{\langle Z/Z_{\odot} \rangle = 0.44\pm0.01}$) compatible with other measurements of similar galaxies in this redshift range. The stellar ages, even if no cosmological constraint is assumed in the fit, show a decreasing trend compatible with a standard cosmological model, proving the robustness of the method in measuring the ageing of the population. Moreover, they show a distinctive mass-downsizing pattern, with more massive galaxies ($\mathrm{\langle\mass\rangle =11.4}$) being older than less massive ones ($\mathrm{\langle\mass\rangle =11.15}$) by $\sim$0.8 Gyr. We thoroughly test the dependence of our results on the assumed star formation history, finding only a maximum 2\% fluctuation on median results using models with significantly different functional forms. The derived ages are combined to build a median age-redshift relation, that we use to perform our cosmological analysis.}
   {By fitting the median age-redshift relation with a flat $\mathrm{\Lambda}$CDM model, assuming a Gaussian prior on $\Omega_{M,0} = 0.3 \pm 0.02$ from late-Universe cosmological probes, we obtain a new estimate of the Hubble constant $\mathrm{H_0 = 67_{-15}^{+14}\: km\: s^{-1}\: Mpc^{-1}}$. In the end, we derive a new estimate of the Hubble parameter by applying the cosmic chronometer method on this sample, deriving a value $\mathrm{H(z = 1.26) = 135 \pm 65\: km\: s^{-1}\: Mpc^{-1}}$ considering both statistical and systematic errors. While the error budget in this analysis is dominated by the scarcity of the sample, this work demonstrates the potential strength of the cosmic chronometers approach up to $z>1$, especially in view of the next incoming large spectroscopic surveys like Euclid.}


   \maketitle
%

\section{Introduction}
\label{sec:1Intro}
Modern Cosmology is nowadays well-established on the successful paradigm of the $\Lambda$CDM model. According to it, dark energy dominates the energy budget of our Universe and is responsible for its accelerated expansion, while cold dark matter (CDM) is the main driver of the gravitational interaction shaping the large-scale structure of the Universe. This description of our Universe was accurately assessed during the last decades thanks to a series of main cosmological probes, like the cosmic microwave background \citep[CMB, e.g.,][]{Bennett2003,PlanckCollaboration2020}, type Ia supernovae \citep[e.g.,][]{Riess1998,Perlmutter1999} or baryonic acoustic oscillations \citep[e.g.,][]{Percival2001,Eisenstein2005}. However, the precision achieved in these measurements has recently revealed a 4-5$\sigma$ tension between the late- and the early-Universe estimates of cosmological parameters \citep{Kamionkowski2022}. It is therefore becoming crucial to extend these measurements beyond the main probes, mapping with independent methods the expansion history of the Universe. This will allow us to understand if the tension is due to some observational or systematic effect, or if the difference is real, hinting towards new physics.

In this context, it is crucial not only to obtain independent measurements of the expansion history of the Universe, but also to find probes that provide such constraints independently of any cosmological assumption. Interest has been growing lately in the analysis of the oldest astrophysical objects, or aiming to determine their absolute ages, setting a lower limit on the current age of the Universe \citep{Valcin2021,Cimatti2023}, or by studying how they age with cosmic time, attempting to trace the evolution of the Universe itself \citep{Moresco2022}.
This second approach is the so-called cosmic chronometers method \citep[CC,][]{Jimenez2002}, which can provide direct estimates of the Hubble parameter from the differential ages of "standard chronometers" with minimal hypotheses. In fact, just assuming the Cosmic Principle and a Friedmann-Lema$\mathrm{\hat{i}}$tre-Robertson-Walker (FLRW) metric, the Hubble parameter can be expressed as:
\vspace{-0.2cm}
\begin{equation}\label{eq:H(z)}
    \mathrm{H(z) = -\frac{1}{1+z} \frac{dz}{dt},}
\end{equation}
meaning that it can be directly measured at a given redshift once the local differential of cosmic age (\textit{dt}) in redshift (\textit{dz}) is known \citep[an extensive description of the method can be found in][]{Moresco2022}.
In order to do this, it is necessary to select a homogeneous and synchronised sample of objects in a certain redshift interval, for which ages or age-related quantities can be measured. In this way, we can obtain an age-redshift that maps the ageing trend of the Universe, and for which Eq. \ref{eq:H(z)} can be applied. 
The best astrophysical objects that can be considered cosmic chronometers over a wide redshift range are massive and passively evolving galaxies. There is a large literature showing that this type of galaxy has built up its mass very rapidly \citep[$\Delta$t<0.3 Gyr,][]{Citro2017,Carnall2018,EstradaCarpenter2019,Carnall2019} and at high redshifts \citep[z>2-3,][]{Thomas2010,Carnall2018}, so that it constitutes a very homogeneous and synchronised population, like chronometers that started ``ticking'' coherently. This is of course an assumption whose impact has to be verified directly on the data. Undoubtedly there is a large literature pointing out ongoing star formation in early-type galaxies, mergers, and continuous formation \citep{Moresco2013,Belli2017} that, if not properly taken into account, might significantly bias our cosmological analysis.

In light of this, an accurate selection process is required to identify, in a sample of galaxies, a group of chronometers, populating the tail of the oldest, most massive objects, and not showing any evidence of ongoing star formation. This requires to analyse different aspects, like the galaxy's colour, its spectral features, or the presence of emission lines and their prominence. In many works \citep{Franzetti2007,Moresco2018,Schreiber2018,Borghi2022a} it was proven how these criteria are indeed complementary and only their combination is able to maximise the purity of the sample.\\
Other than an accurate selection, the CC method requires robust measurements of differential redshift (\textit{dz}) and differential ages (\textit{dt}). If high-precision spectroscopic measurements are today available for the redshift, obtaining accurate age estimates is not as straightforward. Different methods have been explored, some using the whole spectral information, like full-spectral-fitting \citep[e.g.,][]{Carnall2018}, and others focusing on age-related spectral features \citep{Thomas2011}. In \citet{Moresco2011} also another approach was proposed in which, instead of relying on ages, H(z) is determined by tracing the evolution of the break at 4000 {\AA} rest-frame (D4000 hereafter), with the advantage of relying on a direct observable.\\
To date, the CC method has been applied up to redshift z$\simeq$2 adopting both approaches and is very promising in view of upcoming large surveys like Euclid \citep{Laureijs2011}, in which the higher statistics will improve significantly the accuracy on $H(z)$ measurements obtained with this technique.

In this work, we take advantage of the deep VIMOS survey of the CANDELS UDS and CDFS survey fields \citep[VANDELS, ][]{McLure2018} that provides optical spectra and photometry of a wide population of galaxies up to z$\simeq$6.5. It was designed to provide ultra-deep medium-resolution spectra, with enough signal-to-noise ratio (S/N) to perform spectral line studies, both individually on the brighter sources and on stacked spectra for the fainter ones \citep{Garilli2021}. Many works based on VANDELS data have already been published investigating different environments and populations of objects: from the intergalactic medium \citep[e.g.,][]{Thomas2020,Thomas2021} and AGNs \citep[e.g.,][]{Magliocchetti2020}, to Ly$\alpha$ and He II emitters \citep[e.g.,][]{Marchi2019,Hoag2019,Cullen2020,Saxena2020, Saxena2020a,Guaita2020} to the physical properties of star-forming \citep{Cullen2019,Cullen2020,Calabro2021,Calabro2022} and quiescent galaxies \citep{Carnall2019,Carnall2022,Hamadouche2022}. In particular, the last have shown the presence, in VANDELS, of a population of red, massive, and passive galaxies covering the redshift range 1$\leq$z$\leq$1.5, constituting a potential set of chronometers. Moreover, the richness of spectro-photometric information available in VANDELS allows us to adopt a full-spectral-fitting approach to estimate ages as well as many other physical properties of the sample, like metallicity and star formation history. To this purpose, we take advantage of the public code \texttt{BAGPIPES} \citep{Carnall2018}, already tested and validated for VANDELS data, specifically modified as in \citet{Jiao2022} to remove the cosmological prior on ages. By doing so, all the results obtained in this work are independent of any cosmological model.

This paper is divided as follows. In Sect. \ref{sec:2DATA} we present the dataset, the selection process, and the properties of the CCs sample; in Sect. \ref{sec:3METHOD} we describe the full-spectral-fitting process and relative results also discussing the impact of different star formation histories assumptions on the results; in Sect. \ref{sec:4Cosmo} the cosmological analysis is presented, including the $H(z)$ measurement through the CC approach and the assessment of systematic uncertainties; in Sect. \ref{sec:5CONCLUSIONS} we draw our conclusions.


\section{DATA}
\label{sec:2DATA}
In this section we present an overview of the VANDELS survey, the process adopted to select an optimal sample of cosmic chronometers, their spectral features, and physical properties.

\subsection{The VANDELS survey}
\label{sec:2.1TheVANDELSsurvey}
VANDELS \citep{McLure2018,Pentericci2018,Garilli2021} is a deep VIMOS spectroscopic survey targeting high-redshift galaxies in the CANDELS UDS and CDFS survey fields, with a footprint of $\simeq$0.2 $\mathrm{deg^2}$. The observed spectra cover on average the wavelength range $\mathrm{4800\leq\lambda\leq9800}$ {\AA} with a mean spectral resolution R$\simeq$650. Both in UDS and CDFS the CANDELS survey \citep{Grogin2011,Koekemoer2011} offers deep optical-nearIR \textit{HST} imaging and in CDFS also deep \textit{HST}/ACS optical imaging from the GOODS survey \citep{Giavalisco2004} and ultra-deep X-ray imaging \citep{Luo2017}. However, about 50\% of the VANDELS footprint is not covered by \textit{HST} imaging, which lies only in the central areas, and for those objects in the wider-field region, the optical-nearIR photometric information is provided by different ground-based telescopes. A complete list of the available photometric data, covering the wavelength range 3700-45000 {\AA}, can be found in \citet{Garilli2021}.

In this work, we analyse data from the fourth and last data release of VANDELS \citep[DR4,][]{Garilli2021}, which counts 2087 galaxies pre-selected in photometric redshift to lie in the range $\mathrm{1\leq z\leq7}$, comprising 417 star-forming galaxies (SFG, $\mathrm{2.4\leq z\leq5.5}$), 1259 Lyman-break galaxies (LBG, $\mathrm{3.0\leq z\leq7.0}$) and 278 passive galaxies (PASS, $\mathrm{1.0\leq z\leq2.5}$). The remaining 133 objects are AGNs, \textit{Herschel}-detected galaxies, or secondary objects. 
Besides spectra and photometry, DR4 offers also a catalogue including:
\begin{itemize}
    \item spectroscopic redshift measurements ($\mathit{z_{spec}}$) and a relative quality flag ($\mathit{z_{flag}}$);
    \item target classification as one of the types listed above, based on photometric criteria (mainly $i$, $z$, $H$ magnitudes and UV and VJ colours) as described in \citet{McLure2018};
    \item SED-fitting estimates of object properties like: rest-frame UV and VJ colours, stellar mass, V-band dust attenuation, and star formation rate. These quantities are derived using the \texttt{BAGPIPES} code \citep{Carnall2018} as described in \citet{Garilli2021};
    \item correction factors for the error spectra, introduced to improve the correlation between the variance in the observed spectra and the associated median error, as described in Talia et al. (\textit{submitted}).
\end{itemize}

The exposure time for each object (up to 80 hours) is designed to obtain, especially for passive and star-forming galaxies, a S/N high enough to perform detailed spectroscopic studies. This consists in a S/N higher than 10 for star-forming and passive galaxies, and higher than 5 for the other targets \citep{Garilli2021}. 

\subsection{Selecting a reliable sample of cosmic chronometers}
\label{sec:2.2SelectionCosmicChronometers}
A proper application of the cosmic chronometers method requires to select the purest sample of massive and passively evolving galaxies, avoiding contamination by younger star-forming objects that could bias the subsequent cosmological analysis (see \citet{Moresco2022} for more details). Many different criteria have been developed for this purpose, based on rest-frame colours (e.g., UVJ,\citealt{Williams2009}; NUVrJ, \citealt{Ilbert2013}), spectral energy distribution \citep[SED, e.g.,][]{Ilbert2010}, star formation rate \citep[SFR, e.g.,][]{Pozzetti2010} or emission lines \citep[e.g.,][]{Mignoli2009}. In this context, several works have also shown that adopting a simple criterion is not enough to identify all the possible star-forming outliers, with a possible residual contamination of up to 50\%, depending on the criterion \citep{Franzetti2007,Moresco2013}.

\begin{figure*}
     \centering
     \begin{subfigure}[c]{\textwidth}
         \centering
         \includegraphics[width=0.5\textwidth]{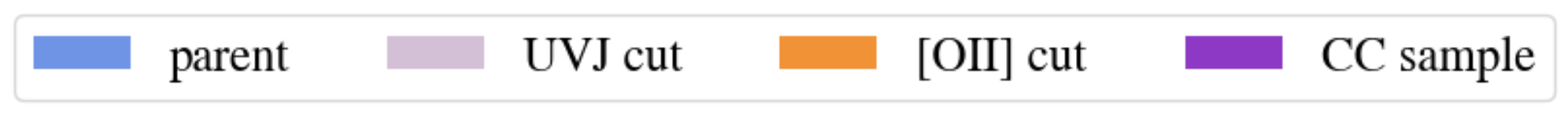}
         \label{fig:histo_leg}
     \end{subfigure}
     \hfill
     \centering
     \begin{subfigure}[b]{0.32\textwidth}
         \centering
         \includegraphics[width=\textwidth]{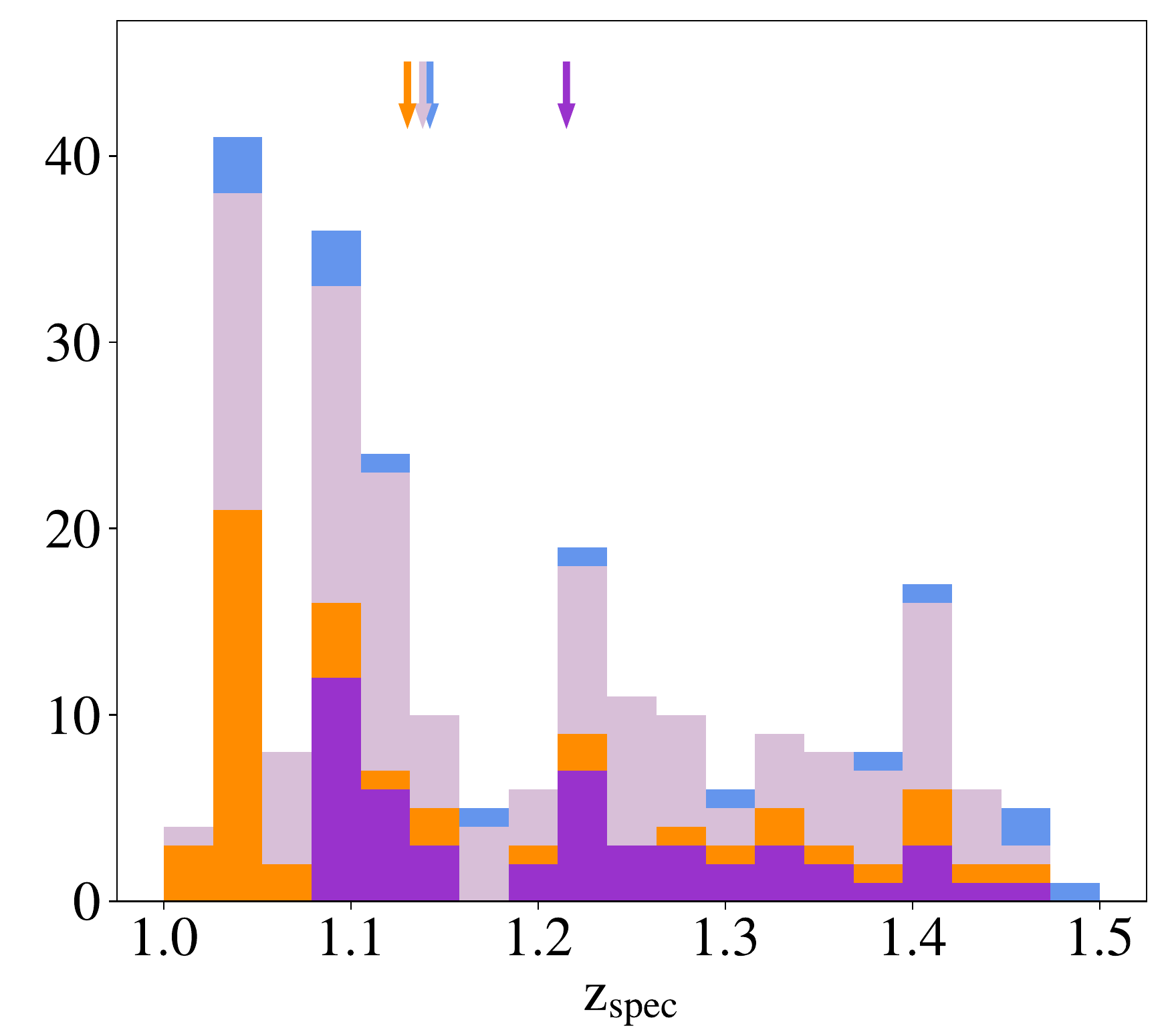}
         \caption{}
         \label{fig:histo_z}
     \end{subfigure}
     \hfill
     \begin{subfigure}[b]{0.32\textwidth}
         \centering
         \includegraphics[width=\textwidth]{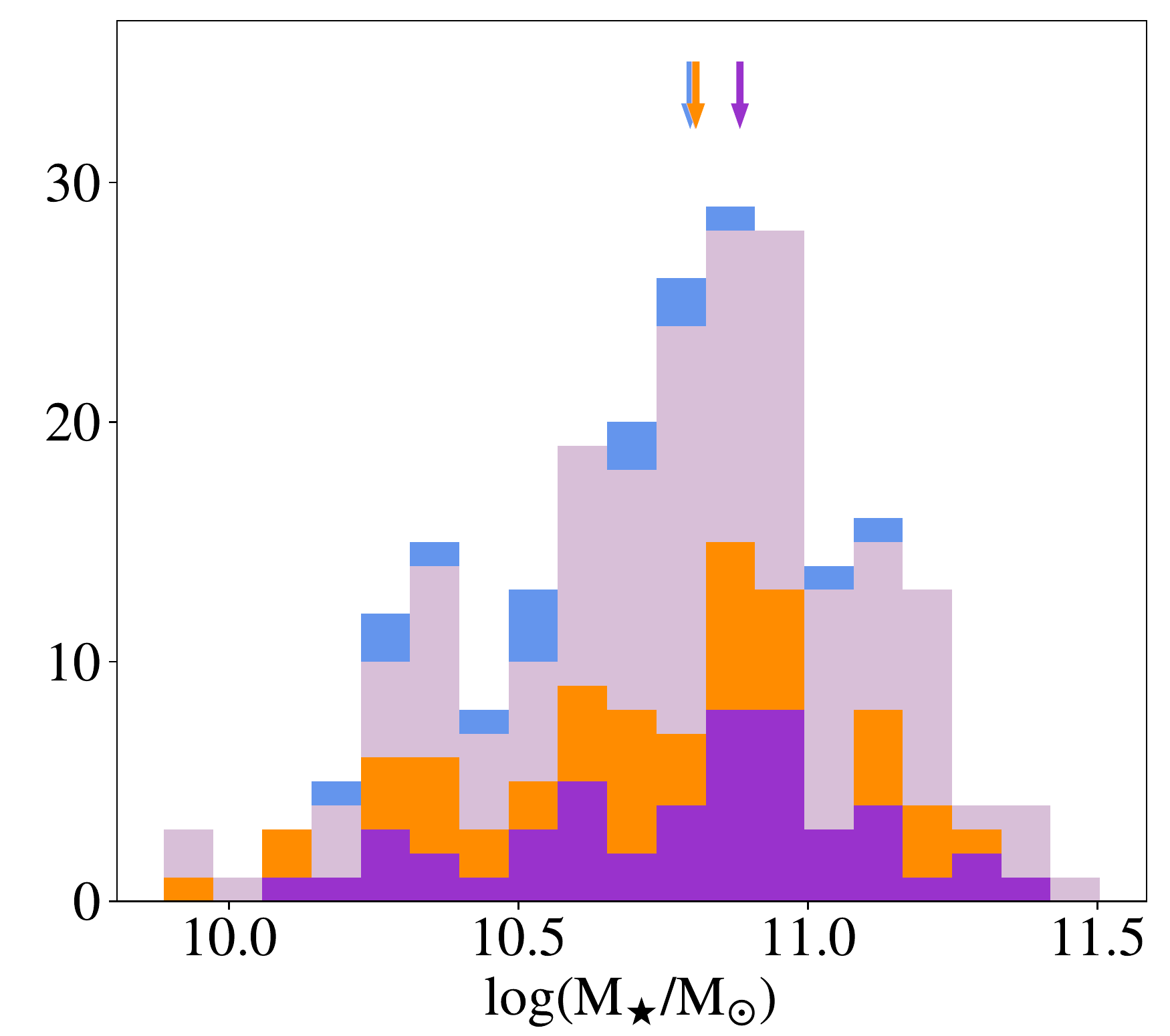}
         \caption{}
         \label{fig:histo_mass}
     \end{subfigure}
     \hfill
     \begin{subfigure}[b]{0.32\textwidth}
         \centering
         \includegraphics[width=\textwidth]{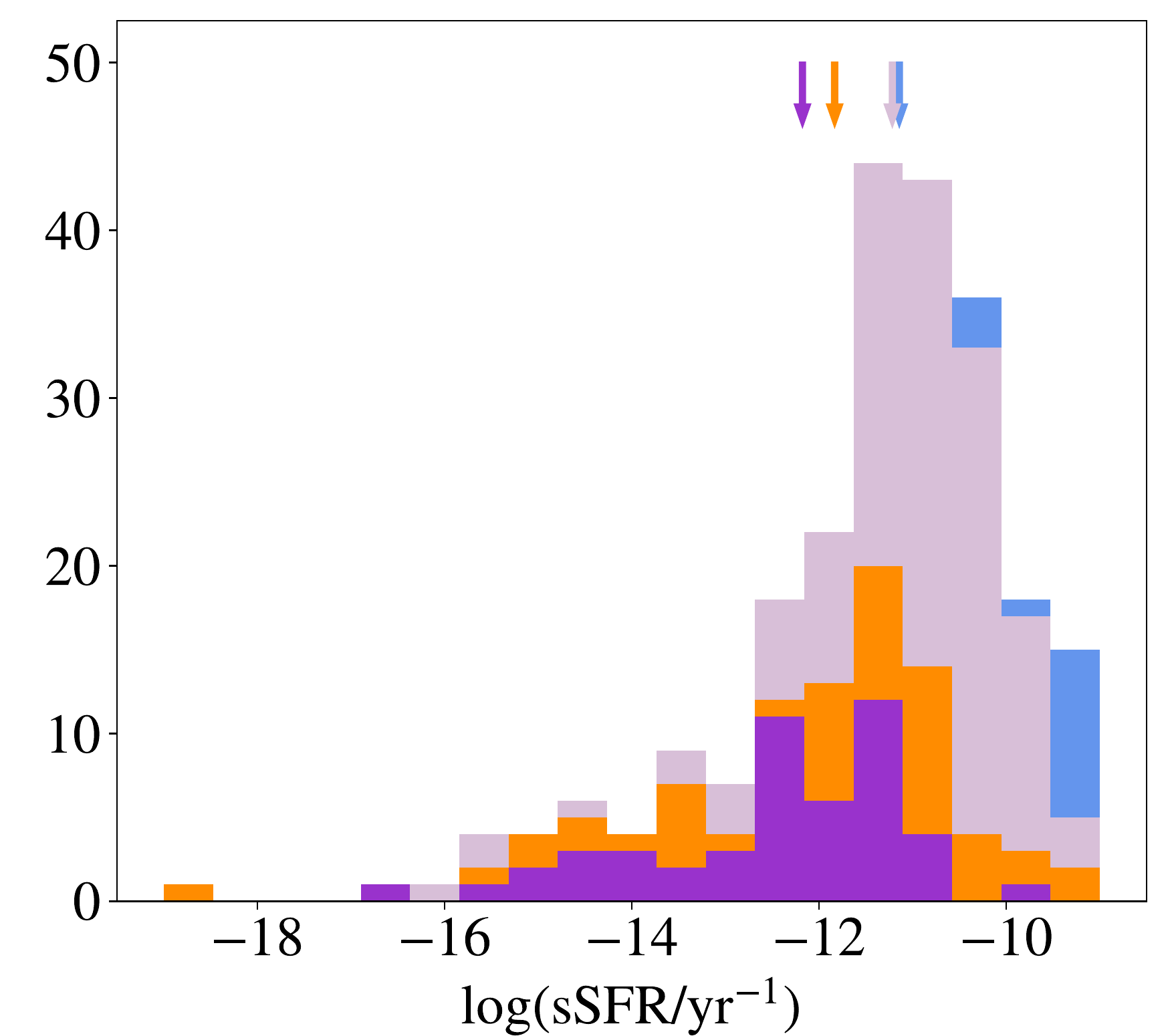}
         \caption{}
         \label{fig:histo_ssfr}
     \end{subfigure}
     \begin{subfigure}[b]{0.32\textwidth}
         \centering
         \includegraphics[width=\textwidth]{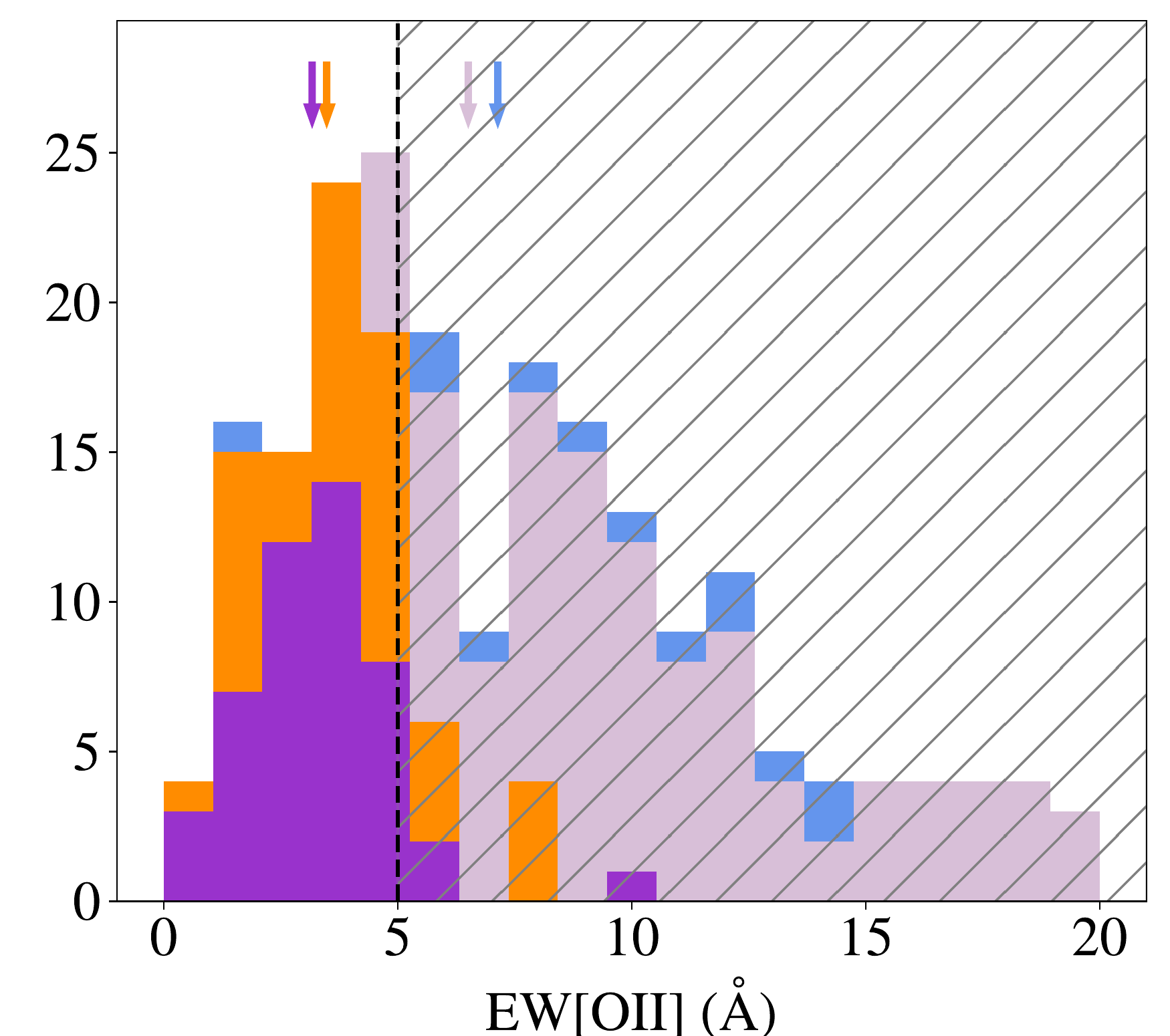}
         \caption{}
         \label{fig:histo_oii}
     \end{subfigure}
     \hfill
     \begin{subfigure}[b]{0.32\textwidth}
         \centering
         \includegraphics[width=\textwidth]{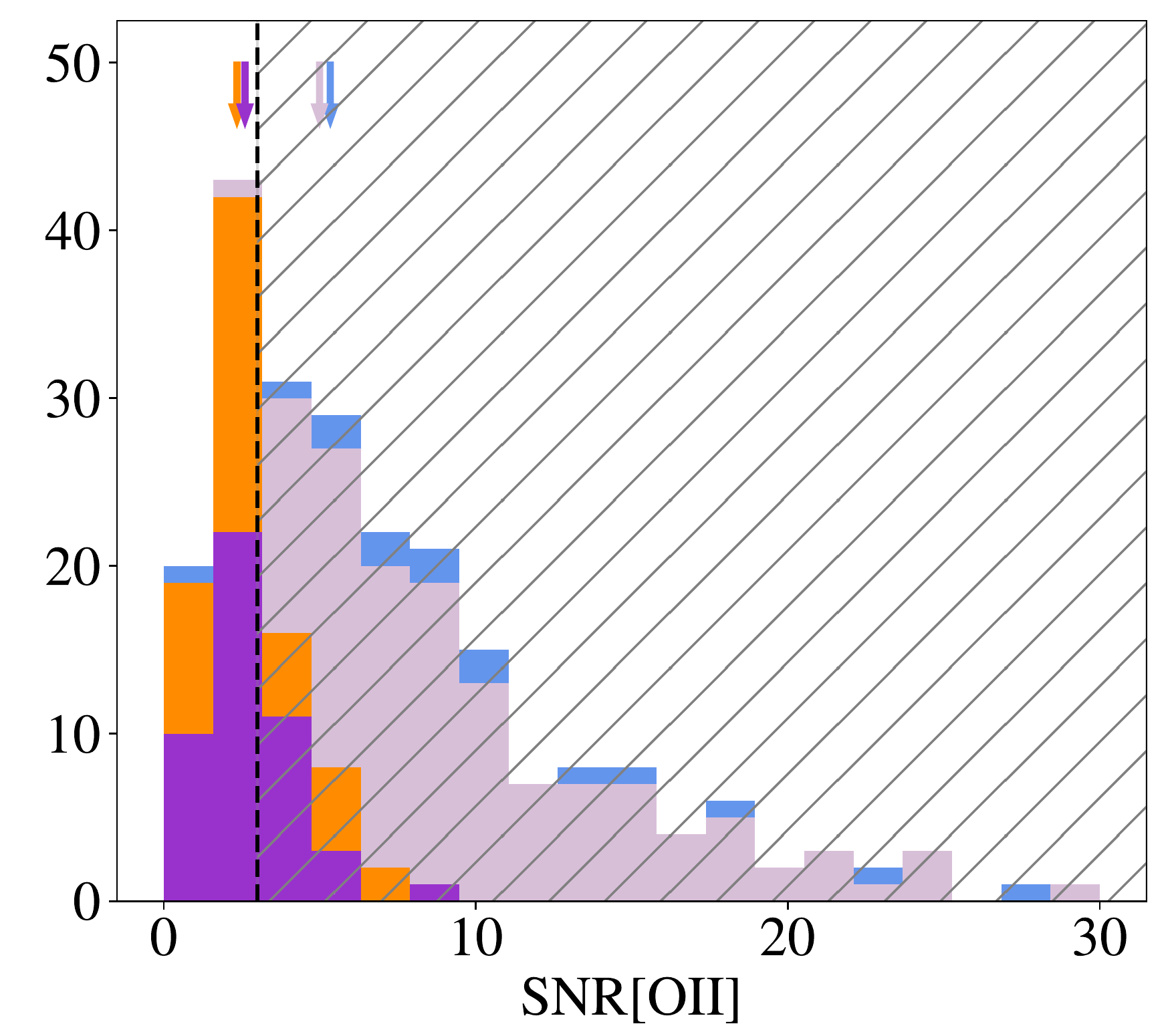}
         \caption{}
         \label{fig:histo_snroii}
     \end{subfigure}
     \hfill
     \begin{subfigure}[b]{0.33\textwidth}
         \centering
         \includegraphics[width=\textwidth]{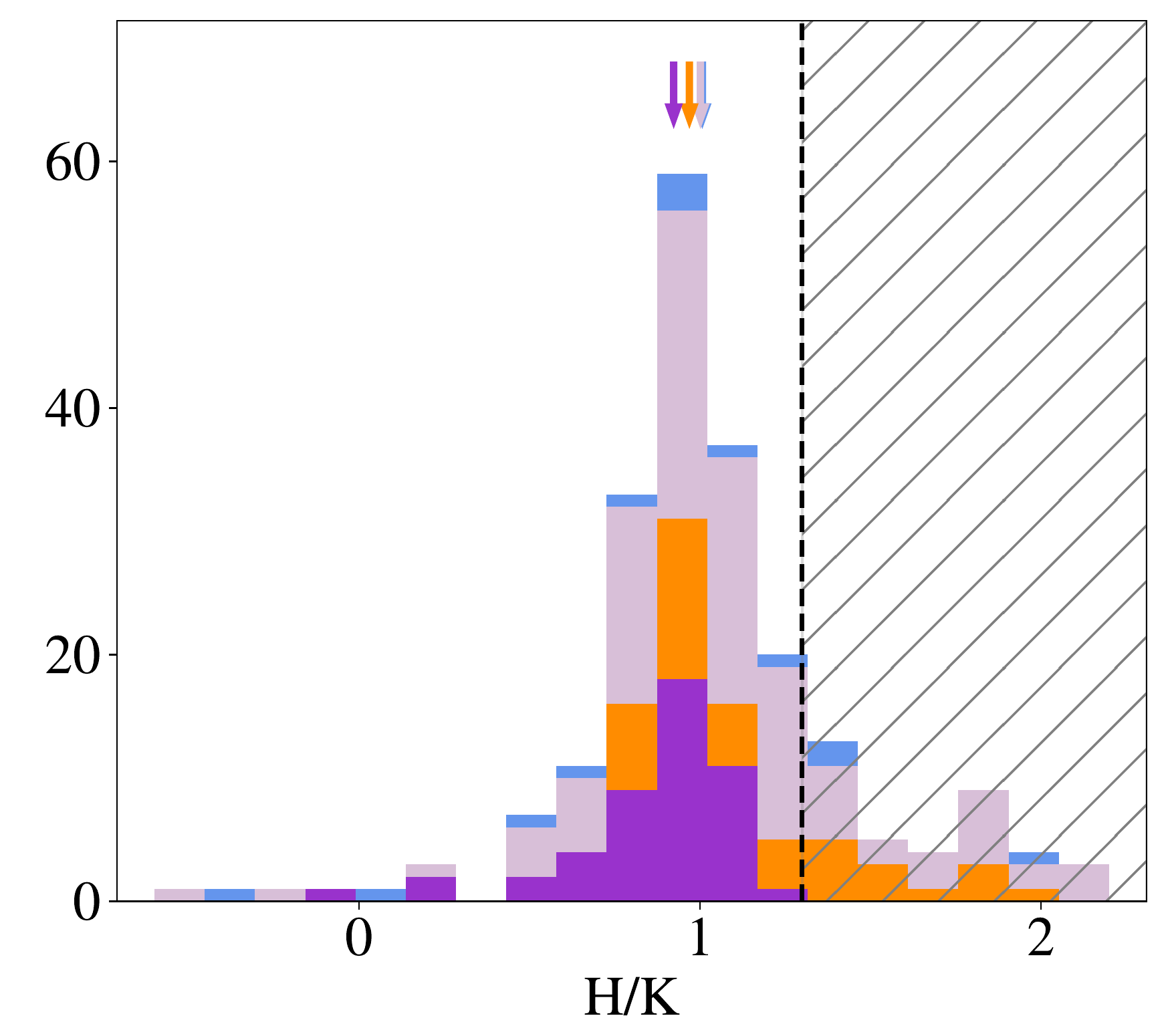}
         \caption{}
         \label{fig:histo_hk}
     \end{subfigure}
        \caption{Histograms for the physical and spectral properties of the samples in each step of selection, as shown in the legend and described in Sect. \ref{sec:2.2SelectionCosmicChronometers}. Vertical arrows represent the corresponding median values and, when present, the dashed vertical line indicates the threshold value adopted to select according to that quantity (the shaded area showing the discarded range). We note that for the [OII] we adopted a conservative cut, discarding only the objects that showed both $\mathrm{EW([OII])>5 \AA}$ and $\mathrm{S/N([OII])>3}$, aiming to
        keep objects in which the [OII] line could be due to just a noise fluctuation.}
        \label{fig:2histograms}
\end{figure*}

For this reason, aiming to maximise the purity of our sample, in this work we combine different and complementary cuts, based on both photometric and spectroscopic information, following this outline:
\begin{enumerate}
    \item \textbf{Parent sample.} As starting point, we define a parent sample made of galaxies selected as follows.
    \begin{itemize}
        \item Classified as passive targets in VANDELS (278 objects). 
        \item In the redshift range 1.0$\leq$z$\leq$1.5. Among the passive targets, 241 galaxies meet this condition, with the lower and upper boundaries of the redshift interval excluding respectively 4 and 33 objects. The lower limit is applied because the distribution of the passive sample becomes statistically significant at z$\geq$1. The upper limit, given the spectra wavelength coverage, is needed to ensure that all spectra include some major spectral features, like the D4000 break, crucial in the following analysis.
        \item With $\mathrm{z_{flag}=3,4}$. In VANDELS, this flag identifies objects with the most reliable redshift measurements, estimated to have >99\% probability of being correct \citep{Garilli2021}. This requirement, met by 265 passive targets, is particularly relevant for the cosmological analysis.
        
        Combining the above criteria, the parent sample counts 234 galaxies.
    \end{itemize}

    \bigskip
    \item \textbf{UVJ cut.} Since the DR4 catalogue provides rest-frame U-V and V-J colours, we apply the photometric criterion based on the colour-colour UVJ diagram \citep{Williams2009}, in order to select photometric passive galaxies, adopting the cut in \citet{McLure2018}:
    \begin{equation}\label{eq:2UVJ}
        \begin{cases}
            \mathrm{U - V > 0.88(V - J) + 0.49}\\
            \mathrm{U - V > 1.2}\\
            \mathrm{V - J < 1.6}
        \end{cases}
    \end{equation}
    As mentioned before, this criterion was already used for the target classification, but the U-V and V-J colours, in that case, were slightly different from those available in DR4 catalogue. In both cases they are derived via SED-fitting but in DR4 this is performed with the additional information of spectroscopic redshifts, an improved photometry and the \texttt{BAGPIPES} code, optimised for this survey. By applying the UVJ cut on the parent sample, 25 additional objects are discarded. Most of them have been identified as post-starburst galaxies in \citet{Carnall2019}. 
    
    After the UVJ selection, our sample reduces to 219 galaxies.

    \bigskip
    \item \textbf{[OII] cut.} We further clean the sample analysing the [OII]$\lambda$3727 emission line, an indicator of ongoing star formation since it is a tracer of photo-ionised gas \citep{MagrisC.2003}. A cut on the equivalent width (EW) of the [OII] is often adopted, like in \citet{Mignoli2009}, where star-forming objects are identified by EW([OII])>5 {\AA}. In this work a more conservative choice is preferred, and only galaxies with a significantly detected [OII] line are discarded, namely objects with EW([OII])>5 {\AA} and S/N([OII])>3. This means that spectra with EW([OII])>5 {\AA} but low S/N are not excluded from the sample, aiming to keep objects in which the [OII] line could be due to just a noise fluctuation.

    The [OII] cut turns out to be the most restrictive selection step, reducing the sample to 96 galaxies (41\% of the parent sample), but is fundamental to minimise the contamination by ongoing star formation or younger components, as discussed in \citet{Moresco2022}.

    \bigskip
    \item \textbf{H/K cut.} Another stellar population diagnostic is the ratio of two absorption lines, CaII H at 3969 {\AA} and CaII K at 3934 {\AA}, first defined by \citet{Rose1984}. In passive galaxies, the CaII K line is generally deeper than the CaII H, but since the CaII H line overlaps with the H$\epsilon$ line of the Balmer series at 3970 {\AA}, deeper in presence of young and hot A and B-type stars, CaII H results more prominent than CaII K if a young component is present in the population. Recent works \citep{Moresco2018,Borghi2022a} have proven the effectiveness of this indicator by showing that already a 5\% contamination by young populations in the flux budget triggers the inversion.

    To quantify this behaviour we evaluate the ratio H/K $\equiv$ (CaII H + H$\epsilon$) / CaII K by measuring the corresponding pseudo-Lick indices with the code \texttt{PyLick} \citep{Borghi2022a}. When using integrated quantities, a common requirement to identify non-contaminated objects is H/K < 1.2-1.5 \citep{Borghi2022a}. In this work, we select galaxies with H/K < 1.3, in order to increase the purity of the sample preserving the statistics, as can be seen in Fig. \ref{fig:histo_hk}.

    The H/K cut reduces the sample to 78 galaxies.

    \bigskip
    \item \textbf{Visual inspection.} The remaining spectra are visually checked to search for anomalies, like residual emission lines or calibration issues. After this passage, the sample counts 74 galaxies.
    
    \bigskip
    \item \textbf{Redshift cut.} On the selected sample we perform a first study of the spectral features, to which we dedicate the next section. This analysis highlights an anomalous behaviour with redshift of some spectral features, making it necessary to add a further step of selection. In particular, we find that all objects below redshift z$<$1.07 present a 4000 $\AA$ break (D4000) weaker than its expected value at those redshifts, with an inconsistent evolutionary trend, which appears to be caused by some systematic effect. Several possible causes for this effect have been explored, and extensively discussed in Appendix \ref{appendixA}, but up to now, no clear evidence is found to account for this anomaly. For this reason, to avoid introducing a potential bias in the subsequent cosmological analysis, we prefer discarding the 23 galaxies below this redshift threshold, obtaining a final sample of 49 cosmic chronometers. We underline that this doesn't affect the robustness of the following results since it only reduces the redshift coverage.
\end{enumerate} 

\begin{table*}
  \centering
  \caption{Median values and associated errors of the most relevant properties describing our sample, according to the different and incremental selection criteria adopted, described in Sect. \ref{sec:2.2SelectionCosmicChronometers}. The names of the samples and relative detailed description can be also found in Sect. \ref{sec:2.2SelectionCosmicChronometers}.}\label{tab:2selection_process}
  \begin{tabular}{lcccc}
  \hline\hline
        & \textbf{parent sample}  & \textbf{UVJ cut} & \textbf{{[}OII{]} cut} & \textbf{CCs sample} \\ \midrule 
                                           
  N° galaxies                              & 234              & 219                 & 96                      & 49                     \\
  \% parent sample                         & 100\%            & 93\%                & 41\%                    & 21\%                   \\
  $\mathrm{z_{spec}}$                      & 1.14 $\pm$ 0.01  & 1.14 $\pm$ 0.01     & 1.13 $\pm$ 0.01         & 1.21 $\pm$ 0.02        \\
  $\mathrm{log(M_\star/M_\odot)}$          & 10.80 $\pm$ 0.02 & 10.80 $\pm$ 0.02    & 10.81 $\pm$ 0.03        & 10.88 $\pm$ 0.05       \\
  $\mathrm{log(sSFR/yr^{-1})}$             & -11.1 $\pm$ 0.1  & -11.2 $\pm$ 0.1     & -11.8 $\pm$ 0.1         & -12.2 $\pm$ 0.2        \\
  S/N{[}OII{]}                             & 5.3 $\pm$ 0.3    & 5.0 $\pm$ 0.3       & 2.3 $\pm$ 0.2           & 2.6 $\pm$ 0.2          \\
  EW{[}OII{]} (\AA)                        & 7.1 $\pm$ 0.3    & 6.5 $\pm$ 0.3       & 3.5 $\pm$ 0.2           & 3.2 $\pm$ 0.2          \\
  H/K                                      & 1.01 $\pm$ 0.02  & 1.01 $\pm$ 0.02     & 0.97 $\pm$ 0.02         & 0.92 $\pm$ 0.02        \\ 
  $\mathrm{S/N_{spec}}$                    & 5.44 $\pm$ 0.11  & 5.46 $\pm$ 0.11     & 5.64 $\pm$ 0.15         & 5.69 $\pm$ 0.17        \\ \bottomrule
  \end{tabular}
\end{table*}
 
\begin{figure}
    \centering
    \begin{subfigure}[t]{0.47\textwidth}
         \centering
         \includegraphics[width=\textwidth]{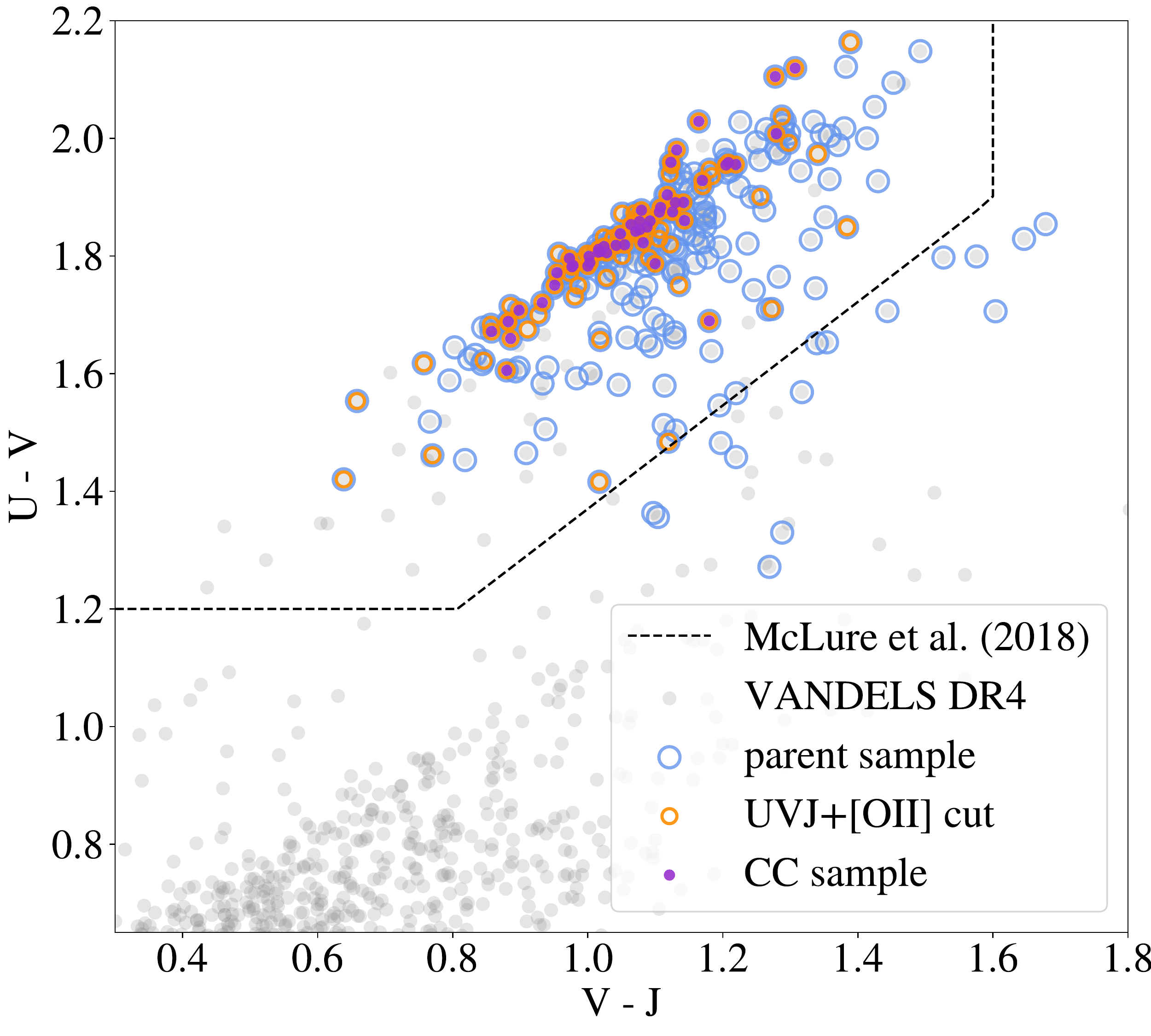}
     \end{subfigure}
     \vspace{-0.5cm}
     \caption{UVJ diagram for the main selection steps. Galaxies above the dashed line are qualified as passive following the criterion in \citet{McLure2018}.}\label{fig:2uvj}
\end{figure}

\begin{figure*}
    \centering
     \begin{subfigure}[t]{0.90\textwidth}
         \centering
         \includegraphics[width=\textwidth]{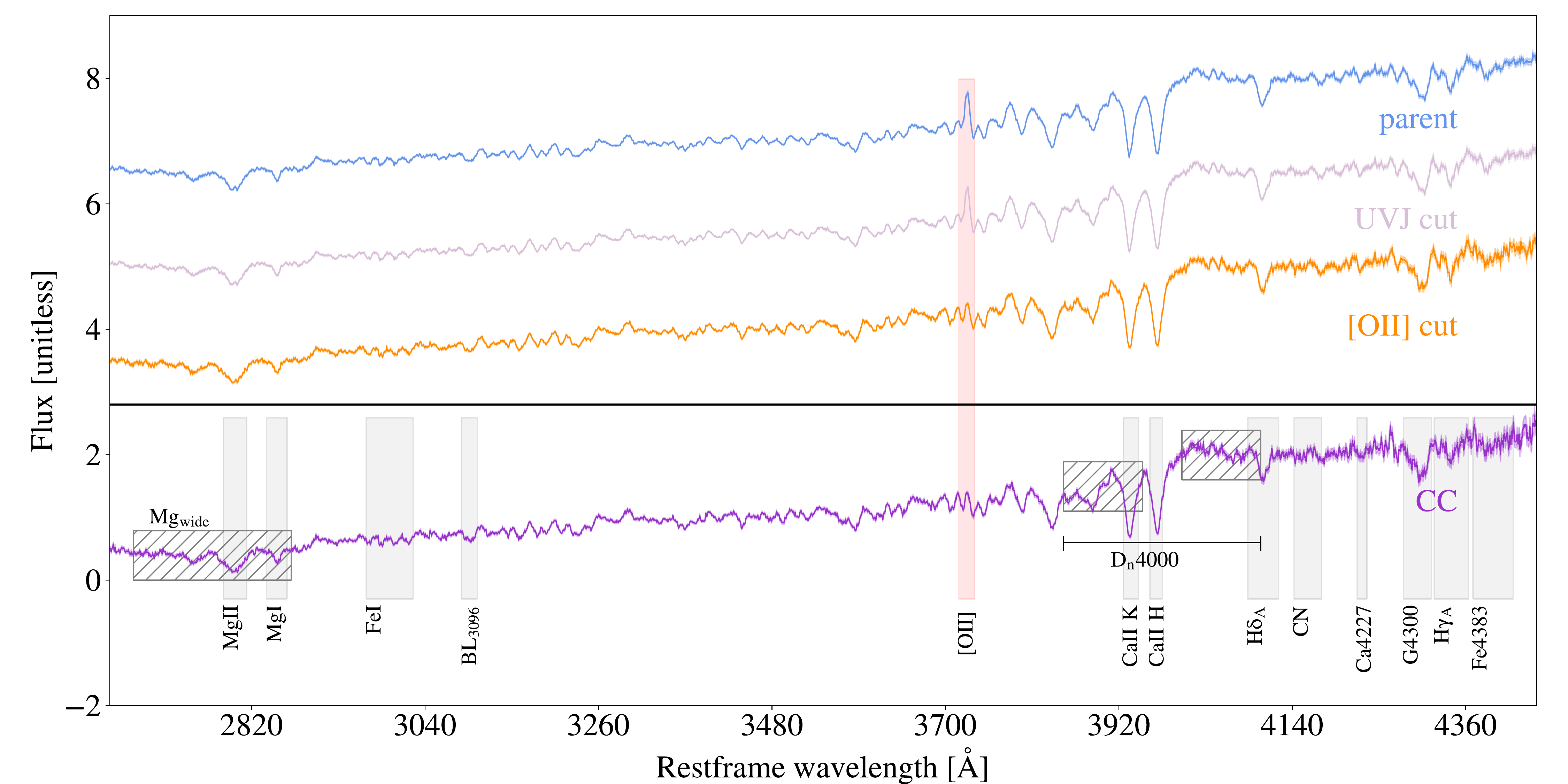}
     \end{subfigure}
     \caption{Median stacked spectra estimated for the different incremental selected samples of our analysis, as described in Sect. \ref{sec:2.2SelectionCosmicChronometers}. We underline that the final sample considered is the purple one, on which we also highlight the most characteristic absorption features (grey shaded area) and spectral breaks (dashed area), as well as the position of potential emission lines (red shaded area), showing the absence of emission lines in our final sample.}\label{fig:2stack}
\end{figure*}

In Tab. \ref{tab:2selection_process}, median values for different physical and spectral properties are listed for each step of the selection process. Unless otherwise specified, errors on median values are computed as median absolute deviations (MAD) divided by the square root of the number of objects. The S/N for each spectrum is estimated as the median of the S/N computed in each pixel in the range $\mathrm{3100 < \lambda < 3500}$ {\AA}. In Fig. \ref{fig:2histograms} is shown the distribution of some of these properties for each passage. With the same colour code, in Fig. \ref{fig:2uvj} can be found the UVJ diagram and in Fig. \ref{fig:2stack} the median stacked spectra for each selection step (normalised in the wavelength range 3320-3850 {\AA}). 

The selected sample of cosmic chronometers, with median redshift $\mathrm{\langle z \rangle = 1.21 \pm 0.02}$, populates the tail of the reddest galaxies in the UVJ diagram and shows the required properties. In particular, it has a median stellar mass equal to $\mathrm{\langle \mass \rangle = 10.88 \pm 0.05}$, and about 75\% of the sample has $\mathrm{\mass > 10.6}$, a value often used as a threshold to select CCs \citep{Moresco2022}. The median specific star formation rate (sSFR) is $\mathrm{\langle \ssfr \rangle = -12.2 \pm 0.2}$, and more than 90\% of the sample has $\mathrm{\ssfr < -11}$, a common limit to characterise passive galaxies \citep{Pozzetti2010}. The histograms in Fig. \ref{fig:2histograms} also show how the selection based on spectral features has been effective in minimising the contamination by young stellar activity, decreasing by 90\% the sSFR obtained after only the UVJ cut. The distributions in [OII] and H/K are typical of passive populations too, with a typical S/N([OII])<3 and a median $\mathrm{\langle H/K \rangle = 0.92 \pm 0.02}$ well below the adopted threshold. In Fig \ref{fig:histo_oii} it can be also noticed the presence of a few objects that, despite having EW[OII]>5 {\AA}, are kept in the CCs sample because of the low S/N[OII].

In Fig. \ref{fig:2stack} the effect of the selection process is shown on median composite spectra. Top-down, the blue stacked spectrum, relative to the parent sample, has the most prominent [OII] emission and very similar CaII H and CaII K. The lilac one, realised after the UVJ cut, is not much different because the sample differs only by 25 objects, mostly post-starburst galaxies \citep{Carnall2019}, but shows a slightly weaker [OII]. Anyway, this confirms again how just adopting a photometric criterion is not enough to remove the contamination by young components, because the [OII] emission line, even if weak, is still present. Only the cut on its EW, which leads to the orange spectrum, is able to clean out this feature. It has a sharp impact on the statistics but is necessary to maximise the purity of the sample. Finally, the purple spectrum is built with the 49 CCs and, besides showing no [OII] emission, has the minimum H/K ratio.

\begin{figure}
     \centering
     \begin{subfigure}[t]{0.42\textwidth}
         \includegraphics[width=\textwidth]{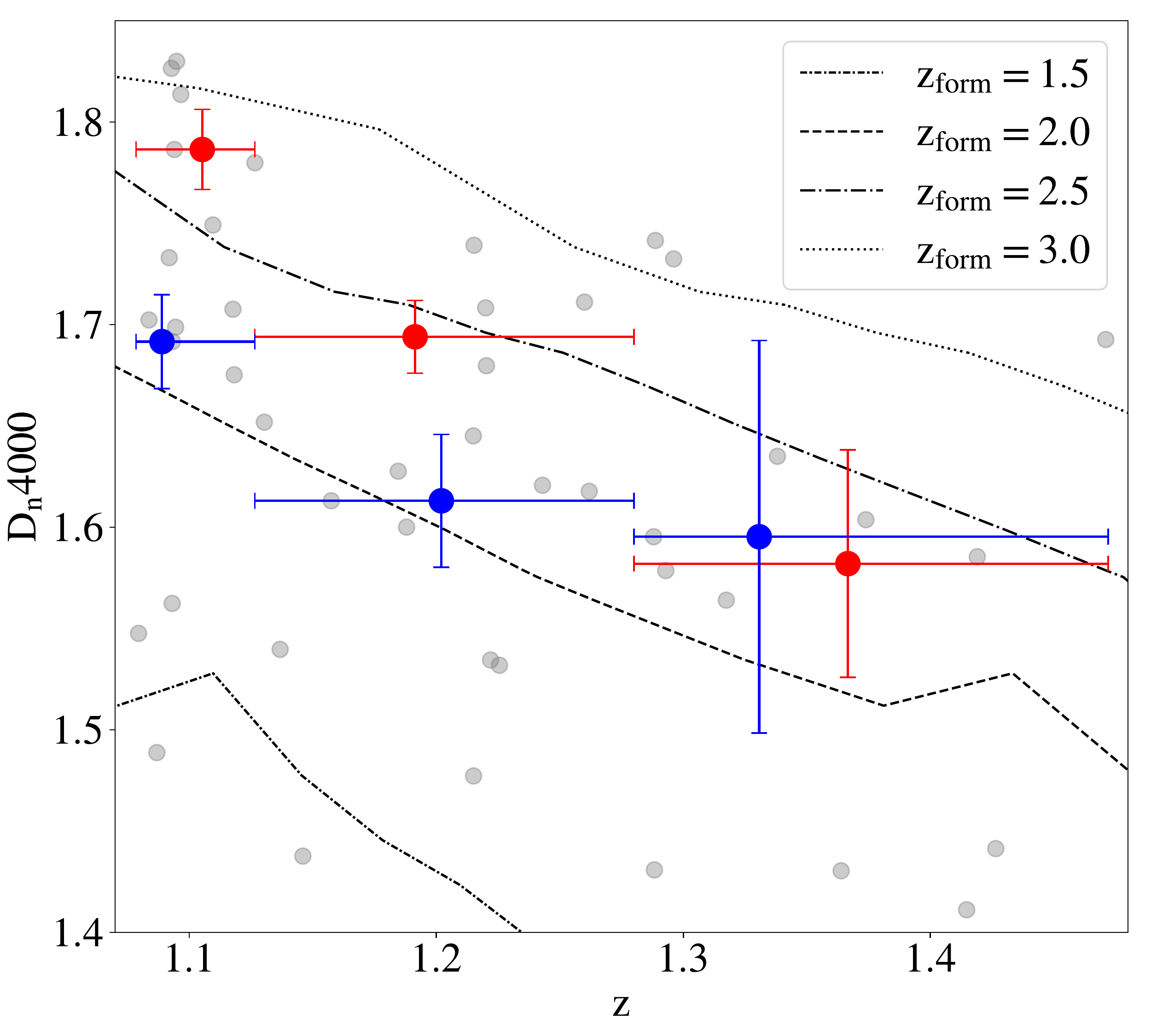}
         \label{fig:2dn4000}
     \end{subfigure}
     \vspace{-0.5cm}
     \caption{$\mathrm{D_n4000}$ trend with redshift. In grey the measurements for the single objects of the CCs sample are shown; blue and red dots are median values averaged in two mass bins, $\mathrm{\mass \leq 10.88}$ (high-mass) and $\mathrm{\mass > 10.88}$ (low-mass) respectively. The different lines show the $\mathrm{D_n4000}$-z relation obtained from BC16 models at different formation redshifts and assuming a reference flat $\Lambda$CDM cosmology ($\mathrm{\lcdm}$), purely for illustrative purposes. We note that qualitatively the observed trends follow quite well the cosmological models, but we defer to Sect. \ref{sec:4Cosmo} the full cosmological analysis.}\label{fig:2D4000}
\end{figure}

\subsection{Spectral properties of the CCs sample}
\label{sec:2.3SpectralProperties}
Since the CCs sample falls in the redshift range 1<z<1.5, the spectral coverage is able to include some interesting spectral features. One of the most relevant is the D4000, a spectral discontinuity at 4000 {\AA} which is particularly strong in the context of passive galaxies. It is caused by a sudden onset of absorption lines at wavelengths bluer than 4000 {\AA} and is stronger in evolved stellar populations. For this reason, at fixed metallicity, it has a tight correlation with the galaxy age, with older galaxies having a stronger D4000, and a very low dependence on the presence of $\alpha$-elements \citep{Moresco2012}. The classical definition of the D4000 is given in \citet{BruzualA.1983}, while a more recent is the \textit{narrow} definition ($\mathrm{D_n4000}$) given in \citet{Balogh1999}, which is less affected by dust absorption. This last one is shown in the purple spectrum in Fig. \ref{fig:2stack}.

As with the CaII H and K, we use \texttt{PyLick} to measure the $\mathrm{D_n4000}$ for the whole sample of CCs. In order to study its trend with redshift and mass, we separate high-mass (HM) and low-mass (LM) galaxies by the median stellar mass of the sample, $\mathrm{10^{10.88} M_\odot}$. We then divide each sub-sample into three redshift bins, such that there is the same number of objects in each bin, getting a total of six sub-samples. In each of them, we measure the median $\mathrm{D_n4000}$ value. In Fig. \ref{fig:2D4000} we show $\mathrm{D_n4000}$ measurements obtained for single galaxies (grey) and median values in each sub-sample, in blue for the LM sample, in red for the HM sample. The $\mathrm{D_n4000}$ shows a clear decreasing trend, both for the HM and the LM sample, giving a first observational evidence that the selected CCs sample ages with cosmic time. We also observe a mass-downsizing trend \citep{Thomas2010}, given that HM galaxies have a stronger median $\mathrm{D_n4000}$ than LM objects, at fixed redshift. These results constitute a first observational validation of the core hypotheses of the CC method.

\section{METHOD AND ANALYSIS}
\label{sec:3METHOD}
In this section, we present the method adopted to estimate ages and physical properties of the CCs sample, the code used, its settings, and the results obtained.

\subsection{Full-spectral-fitting with \texttt{BAGPIPES}}
\label{sec:3.1FSFwithBagpipes}
To these days there are two main ways of estimating ages from galaxy spectra: by fitting observed spectra with synthetic ones, and so using the whole spectral information, or by focusing on particular age-dependent indices. In this work, to benefit from the richness of the spectral and photometric information available in VANDELS, we follow the first path. In particular, we exploit a full-spectral-fitting technique using the public code \texttt{BAGPIPES} \citep{Carnall2018}, already employed and optimised on VANDELS data, which allows us to fit jointly spectra and photometry with a Bayesian approach. This means that once chosen a model $\mathcal{M}$ depending on a set of parameters $\theta$, the observed spectrum $\mathrm{f_\lambda}$ is modelled maximising the posterior probability on $\theta$, $\mathcal{P}(\theta | f_\lambda, \mathcal{M})$, defined by the Bayes theorem:
\begin{equation}
    \label{eq:3posterior}
    \mathcal{P}(\theta | f_\lambda, \mathcal{M}) = \frac{\mathcal{L}(f_\lambda|\theta, \mathcal{M})\;\mathcal{P}(\theta | \mathcal{M})}{\mathcal{P}(f_\lambda, \mathcal{M})}
\end{equation}
and sampled through the nested sampling algorithm \texttt{MultiNest} \citep{Buchner2014}. In particular, as extensively described in \citet{Carnall2019}, the model is built with four main components:
\begin{itemize}
    \item SSP($\lambda$, age, Z), a simple stellar population model depending on the wavelength $\lambda$, on the age of the stellar population, on its metallicity Z, and on the initial mass function (IMF). The stellar population synthesis (SPS) models, implemented in the form of grids of SSPs, are the 2016 version of \citet{Bruzual2003} \citep[see][]{Chevallard2016}. They are constructed using a \citet{Kroupa2002} IMF;
    \item SFR(t), the galaxy star formation history (SFH). The code allows us to combine different SFHs, one for each SSP, with various functional forms (e.g. delta function, constant, exponentially declining, double-power-law);
    \item $\mathrm{T^+(age,\lambda)}$, transmission curve of the ionised interstellar medium (ISM) as described in \citet{Charlot2001}. This component is referred to as nebular and, unlike the ones above, is optional;
    \item $\mathrm{T^0(age,\lambda)}$ transmission curve of the neutral ISM, mainly due to dust absorption and emission. Different models are implemented in the code, like \citet{Calzetti2000}, \citet{Cardelli1989}, or \citet{Charlot2000}. This is an optional component too and is referred to as dust.
\end{itemize}
In addition to these, there are two more optional but non-physical components: noise, which is an additive term acting on the error spectrum to correct a possible underestimation, and calibration, which perturbs the spectrum with a second-order Chebyshev polynomial to fix possible calibration issues.\\
The code provides, for each galaxy, a best-fit spectrum and an estimate to the $\mathrm{16^{th},\: 50^{th},\: and\: 84^{th}}$ percentile of all the parameters involved in the fit (e.g., age, metallicity, and stellar mass formed). The $\mathrm{M_{formed}}$ parameter is defined as the whole stellar mass formed from t=0 to the time $t$ at which it is observed:
\begin{equation}
    \mathrm{M_{formed} = \int_0^t SFR(t')dt'}.
\end{equation}
and in the following we will refer to this parameter when discussing the mass of the galaxies.
This parameter is slightly different from the standard stellar mass, which does not include stellar remnants, but the two are tightly correlated. In general, we find an offset of $\sim$0.2 dex between the two, that we need to take into account when comparing our resulting masses with literature values.

In this context, it is important to underline that the version of \texttt{BAGPIPES} we are using differs from the original for the treatment of the prior on galaxies' ages. This modification is described in detail in \citet{Jiao2022}. Originally the code assumes a cosmological prior on ages, for which the maximum age of the fit should be smaller than the age of the Universe at the corresponding redshift, given a flat $\Lambda$CDM model with parameters $\mathrm{\Omega_{M}=0.3,\: \Omega_\Lambda=0.7\: and\: H_0=70\: km\:s^{-1}Mpc^{-1}}$. Although its impact is of relative interest in galaxy evolution studies, and is generally neglected, it can't be ignored in cosmological analyses because the derived ages would be constrained by the cosmological model assumed, and hence fitting those one would just retrieve the assumed prior. To avoid this, we consider instead a uniform prior on galaxy ages, so that they can span between 0 and 20 Gyr independently of redshift. This implies that, in theory, the estimated ages could result greater than the age of the Universe as expected in a standard cosmological model at a given redshift, so a first test will be checking their compatibility. 
\subsubsection{SFH functional form}
\texttt{BAGPIPES} is able to combine multiple SFHs, one for each stellar population, with different functional forms. We present here two of the most extensively used, on which we will focus our analysis:
\begin{itemize}
        \item the \textbf{delayed exponentially declining (DED)} SFH, given by the equation
        \begin{equation}\label{eq:3delayedSFH}
            \mathrm{SFR(t) \propto \begin{cases} (t-T_0) e^{-\frac{t-T_0}{\tau}}, & t > T_0 \\ 0, & t < T_0 \end{cases}},
        \end{equation}
        where $\tau$ sets the width of the SFH while $\mathrm{T_0}$ is the age of the Universe when the star formation begins. This means that it is directly linked to the age of the galaxy, which can be obtained by:
        \begin{equation}\label{eq:3age}
            \mathrm{age = age_U(z_{obs}) - T_0},
        \end{equation}
        where $\mathrm{age_U(z_{obs})}$ is the age of the Universe at the redshift of observation. Besides this, \texttt{BAGPIPES} provides also another age definition, called \textit{mass-weighted age}:
        \begin{equation}\label{eq:3age_MW}
          \mathrm{age_{mw} = age_U(z_{obs}) - \frac{\int_0^{t_{obs}} tSFR(t)dt}{\int_0^{t_{obs}} SFR(t)dt}}.
         \end{equation}
        This is similar to Eq. \ref{eq:3age} but $\mathrm{T_0}$ is replaced with an estimate of the age weighted on the galaxy SFR. In the case of a DED SFH the results obtained with these two definitions differ by an offset ($\mathrm{\sim 0.6\: Gyr}$), typically constant in redshift and age.\\
        This SFH is frequently used \citep{Citro2017,Carnall2018} for passively evolving galaxies, characterised by a single and strong episode of star formation followed by passive evolution, because it is able to reproduce a realistic star formation process using only two parameters;
        \item the \textbf{double-power-law (DPL)} SFH, given by:
        \begin{equation}\label{eq:3dplSFH}
        \mathrm{SFR(t)} \propto \left[\left(\frac{t}{\tau}\right)^\alpha + \left(\frac{t}{\tau}\right)^{-\beta}\right]^{-1},
      \end{equation}
      where $\alpha$ and $\beta$ describe respectively the falling and rising slope of the curve while the $\tau$ parameter is related to the SFH peak. In this case, differently from a DED SFH, due to the shape of the SFH, the only age definition provided is the mass-weighted one. This type of functional form allows the SFH more freedom in shape by decoupling the rising and falling phase of the star formation, but the increased number of free parameter might be carefully studied since it might induce non-physical correlation between those (see App. \ref{appendixB}).
    \end{itemize}
    
\subsection{Full-spectral-fitting in VANDELS}
\label{sec:3.2FSFinVANDELS}
Before performing the fit on VANDELS data, we check spectra and photometry to fix possible anomalies. On the one hand, we define a spectroscopic S/N by dividing at each wavelength the flux for its associated noise which allows us to determine the average S/N of the spectrum. On the other hand, analysing the photometric S/N distribution, we observe that it is up to three orders of magnitude higher (max($\mathrm{S/N_{phot}) \sim 10^3}$) than the spectroscopic one ($\mathrm{\langle S/N_{spec} \rangle = 5.74 \pm 0.17}$) for the same object, especially at the reddest wavelengths. If unaccounted for, this difference can have a significant impact on the fit, forcing it to reproduce the photometry irrespectively of the spectroscopic data. We also observe that the redder photometric points are also the ones with smaller errors, and this could introduce an additional bias in the analysis since the bluer photometric data points have been proven to be also the fundamental ones in reconstructing the physical properties linked to the star formation activity. To correct for this issue, we assume a maximum S/N for the photometric data points to S/N=10, correcting the errors in case it was larger and in this way better weighting all the parts of spectrum and photometry.

Once performed these corrections, we test different configurations of parameters and priors, with a particular focus on the impact of the chosen SFH. This is important not only to accurately reproduce the observed data but also to evaluate how much the results are robust when changing the fitting model, which is fundamental also to estimate potential biases and systematic effects when performing cosmological analyses. Here we report parameters and priors of three fit settings: baseline, configuration 1, and configuration 2, as reported in Tab. \ref{tab:3priors}. They differ mainly in two characteristics: the SFH functional form and the data used in the fitting process. The baseline and config. 1 models both fit spectra and photometry but with a different SFH, respectively a DED and a DPL; the baseline and config. 2 models, instead, both use a DED SFH but the second one does not include photometry in the fit. Apart from these differences, all these three models are built on a common set of components, namely: 
\begin{itemize}
    \item a SFH modelled either with a DED or a DPL functional form;
    \item a dust attenuation component described in \citet{Salim2018}, which consists of a power law like in \citet{Calzetti2000} but with a deviation from the slope parameterised by $\delta$;
    \item a nebular component, implemented in \texttt{BAGPIPES} using the code \texttt{Cloudy} \citep{Ferland2017} and described in more detail in \citet{Carnall2018}. The process of selection should have already excluded objects having this type of emission, but we conservatively decide to include it in the model as an additional check of its absence;
    \item redshift fixed, for each galaxy, to the value of the spectroscopic redshift obtained in VANDELS;
    \item a calibration component, implemented as a second-order Chebyshev polynomial;
    \item a noise correction, introduced as white noise.
\end{itemize}
The main parameters and relative priors are listed in Tab. \ref{tab:3priors}.

\begin{figure*}
    \centering
    \includegraphics[width=0.98\textwidth]{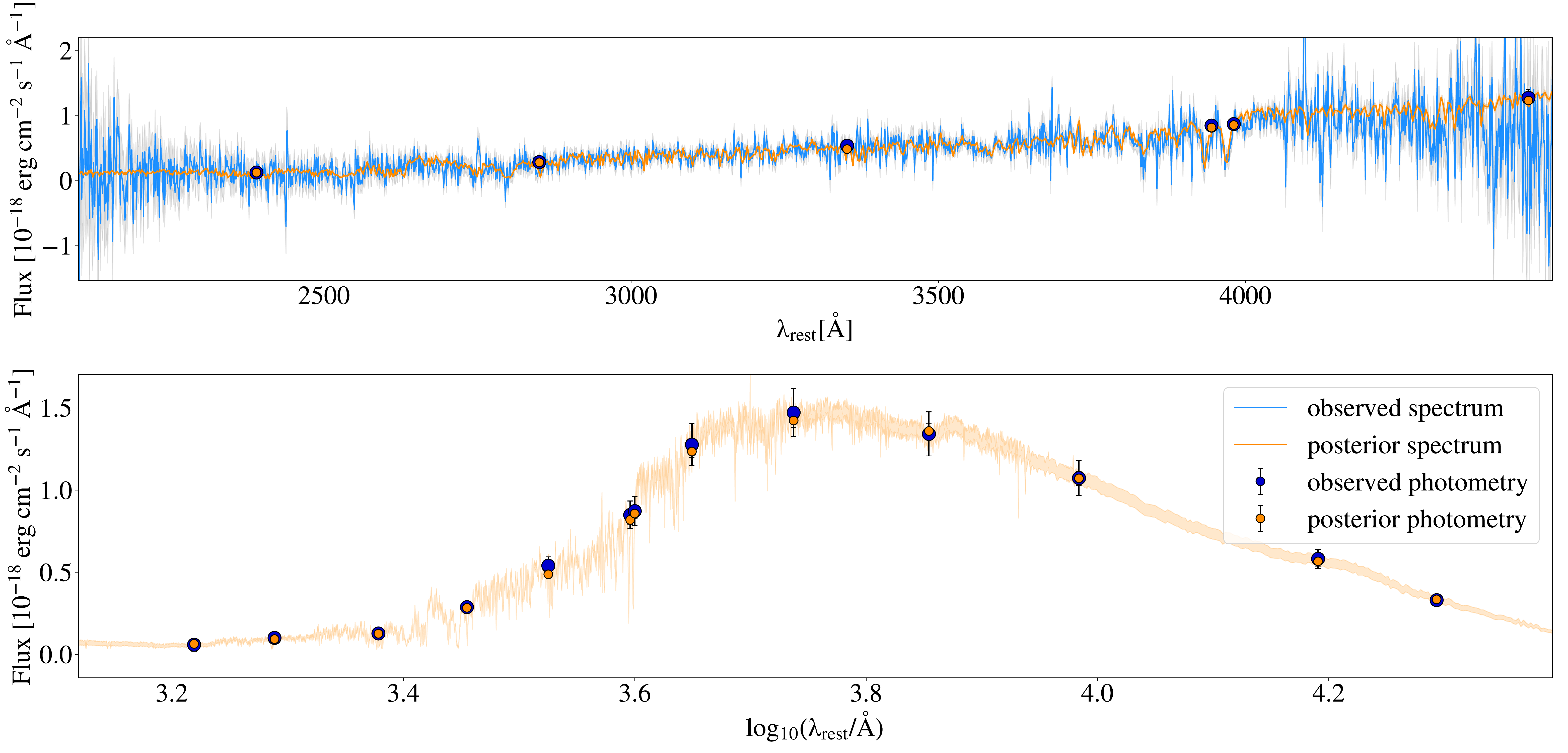}
    \caption{Example of best-fit spectrum and photometry (in orange) obtained by fitting the observed ones (in blue) with the baseline configuration, as described in Sect. \ref{sec:3.2FSFinVANDELS}.}
    \label{fig:3post_spec_phot}
\end{figure*}

For each configuration, we verify the convergence of the results and that the best-fit model is correctly reproducing the observed spectra and photometry. If these requirements aren't met, the results can't be considered valid, and are therefore flagged as bad fit. To perform this operation, we examine visually the best-fit spectra and photometry and the distribution of the posterior probability on the parameters for each galaxy. For the baseline configuration, only 5/49 galaxies are flagged as bad fit. In Fig. \ref{fig:3post_spec_phot} the typical best-fit spectrum and photometry are shown in the case of a good fit for the baseline configuration. 

As the name implies, the baseline configuration will be used as a benchmark to perform both the analysis of the physical properties of the population and the cosmological study, which will be presented in the next sections. Among the three configurations presented here, indeed, it presents the smaller number of bad fits and the best agreement between observed spectrophotometry and the posterior one. In Appendix \ref{appendixB} we instead discuss the comparison between baseline and config. 1 results, aiming to evaluate how the choice of a more complex SFH impacts the results. This also allows us to account for its systematic effect in the cosmological analysis. 

\begin{table*}
\centering
\caption{Parameters and priors adopted for each configuration used in the full-spectrum-fitting. Columns 1-2 indicate if spectrum and photometry are considered (\cmark) or not (\xmark) in the fit. Columns 3-6 show the type of prior and the range for some main parameters. Columns 7-8 report the type of star formation history and relative parameters. The acronyms DED and DPL refer respectively to the delayed exponentially declining and double power law SFHs, while the symbol $\mathcal{U}$ indicates a uniform prior with the associated range.}
\label{tab:3priors}
\resizebox{\textwidth}{!}{%
\begin{tabular}{@{}lcccccccc@{}}
\hline\hline
         & 
spectrum & 
photometry & 
$\mathrm{log(\sigma_{vel}/km\:s^{-1})}$& 
age {[}Gyr{]} & 
$\mathrm{\massf}$ & 
$\mathrm{log(Z/Z_\odot)}$ & 
SFH &
SFH parameters\\ 

        &
(1)     &
(2)     &
(3)     &
(4)     &
(5)     &
(6)     &
(7)     &
(8)     \\

\midrule
baseline & 
\cmark & 
\cmark & 
$\mathcal{U}(1, 2.7)$ & 
$\mathcal{U}(0, 20)$   & 
$\mathcal{U}(0,13)$ & 
$\mathcal{U}(-0.85,0.24)$ & 
DED & 
$\mathrm{\tau \in \mathcal{U}(0,1)}$ \\

config. 1 & 
\cmark          & 
\cmark        & 
$\mathcal{U}(1, 2.7)$& 
$\mathcal{U}(0, 20)$   & 
$\mathcal{U}(0,13)$ & 
$\mathcal{U}(-0.85,0.24)$ & 
DPL & 
\begin{tabular}[c]{@{}l@{}} $\mathrm{\tau \in \mathcal{U}(0,20)}$\\$\mathrm{log(\alpha) \in \mathcal{U}(-2, 3)}$\\$\mathrm{log(\beta) \in \mathcal{U}(-2, 3)}$\end{tabular} \\

config. 2 & 
\cmark & 
\xmark & 
$\mathcal{U}(1, 2.7)$ & 
$\mathcal{U}(0, 20)$ & 
$\mathcal{U}(0,13)$ & 
$\mathcal{U}(-0.85,0.24)$ & 
DED & 
$\mathrm{\tau \in \mathcal{U}(0,1)}$\\ 
\bottomrule
\end{tabular}%
}
\end{table*}

\subsection{Physical properties of the CCs sample}\label{sec:3.3PhysProp}
The fitting process with the baseline configuration is successful for 44 CCs (90\% of CCs sample), with a median reduced chi-square of $\Tilde{\chi}^2$=1.46, considering both spectra and photometry. In order to estimate their physical properties, we compute, for each parameter, the median and $\mathrm{16^{th}}$-$\mathrm{84^{th}}$ percentile ranges of the posterior distribution. The analysis of these quantities shows that our CCs sample, in agreement with our selection process, consists of galaxies with:
\begin{itemize}
    \item high stellar masses, as commonly occurs in a population of passive galaxies, with a median stellar mass $\mathrm{\langle \massf \rangle}$ $\mathrm{=11.21 \pm 0.05}$; 
    \item a median metallicity value of $\mathrm{\langle Z/Z_\odot \rangle = 0.44 \pm 0.01}$; 
    \item short SFHs, with a typical timescale for the star formation $\mathrm{\langle \tau \rangle = 0.28 \pm 0.02}$ Gyr. This also agrees with a mass-downsizing scenario,  where more massive galaxies are the first to assembly in very short bursts of formation \citep{Thomas2010,Citro2017};
    \item low dust reddening, with a median V-band dust extinction of $\mathrm{\langle A_{V,dust} \rangle = 0.43 \pm 0.02}$ mag.
\end{itemize}
We notice here that in the literature there is quite a large spread in the derived metallicity for passive and massive galaxies at these redshifts, with estimates going from sub-solar values 0.4-0.7 $\mathrm{Z_\odot}$ \citep[e.g., ][]{Kriek2019,Lonoce2020,Carnall2022} up to 1.3-1.6 $\mathrm{Z_\odot}$ \citep[e.g.,][]{Conroy2014,Onodera2015}. Our results, in particular, are compatible with the ones found in \citet{Kriek2019} and \citet{Lonoce2020}, and slightly lower with respect to the metallicities obtained by \citet{Carnall2022}, who analysed the stacked spectra from a sample of VANDELS passive galaxies in the range $\mathrm{1\leq z\leq 1.3}$, finding 
$\mathrm{Z/Z_\odot = 0.74^{+0.15}_{-0.12}}$ with \texttt{BAGPIPES}.\\
We also compare our results with the independent analysis of \citet{Saracco2023}, who studied a sample of 64 passive galaxies in VANDELS performing non-parametric full-spectral-fitting on stacked spectra in six mass ranges, using different combinations of models and fitting codes. The sample was selected following the same requirements adopted in this work to identify the parent sample but with a tighter redshift range ($\mathrm{1\leq z\leq 1.4}$) and adding a lower limit to the S/N (S/N>6 per {\AA} in the range [3400-3600] {\AA}). They find typically over-solar metallicities, different from both the ones in this work and in \citet{Carnall2022}. The spread in these results could be due to the models adopted or to the full-spectrum-fitting code used, as also shown in \citet{Saracco2023}, where the variation of one or both of them produces a scatter in the metallicity estimation of up to 0.4 dex.\\ 
It is important to underline that, for the cosmological purpose of this work, we expect the impact of this effect to be negligible, given that the CC method relies on the estimation of differential quantities. In particular, given an optimally selected sample of massive and passive galaxies, what is fundamental is to ensure its homogeneity. This condition is met in our sample, that shows homogeneous metallicities in the considered redshift range. Moreover, comparing our age estimates with the ones in \citet{Saracco2023}, we find a good agreement with their mass-weighted ages, both using the code \texttt{STARLIGHT} \citep{CidFernandes2005} and \texttt{pPXF} \citep{Cappellari2017}.

\begin{figure*}
    \centering
     \begin{subfigure}[c]{\textwidth}
         \centering
         \includegraphics[width=0.4\textwidth]{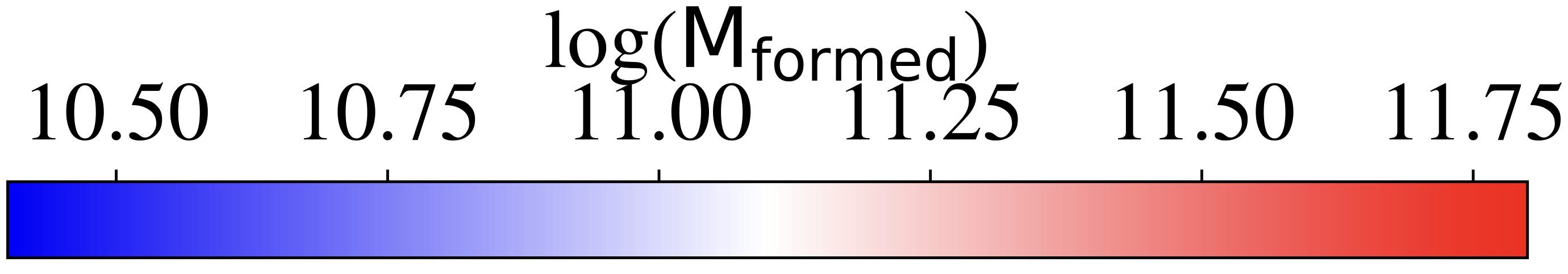}
         \label{fig:legend}
     \end{subfigure}
     \hfill
     \centering
     \begin{subfigure}[b]{0.5\textwidth}
         \centering
         \includegraphics[width=0.9\textwidth]{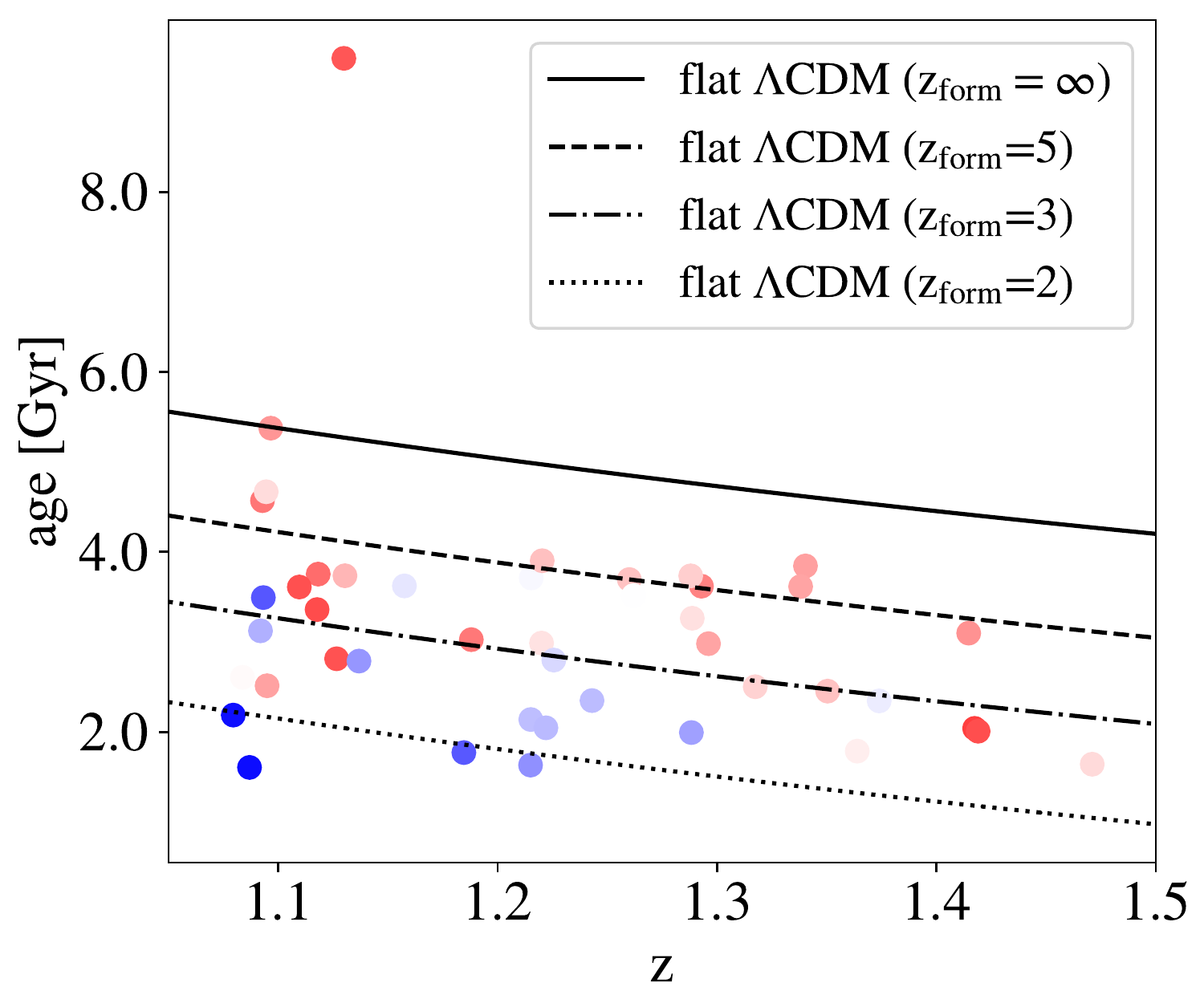}
         \caption{}
         \label{fig:age_z}
     \end{subfigure}
     \hfill
     \begin{subfigure}[b]{0.49\textwidth}
         \centering
         \includegraphics[width=0.9\textwidth]{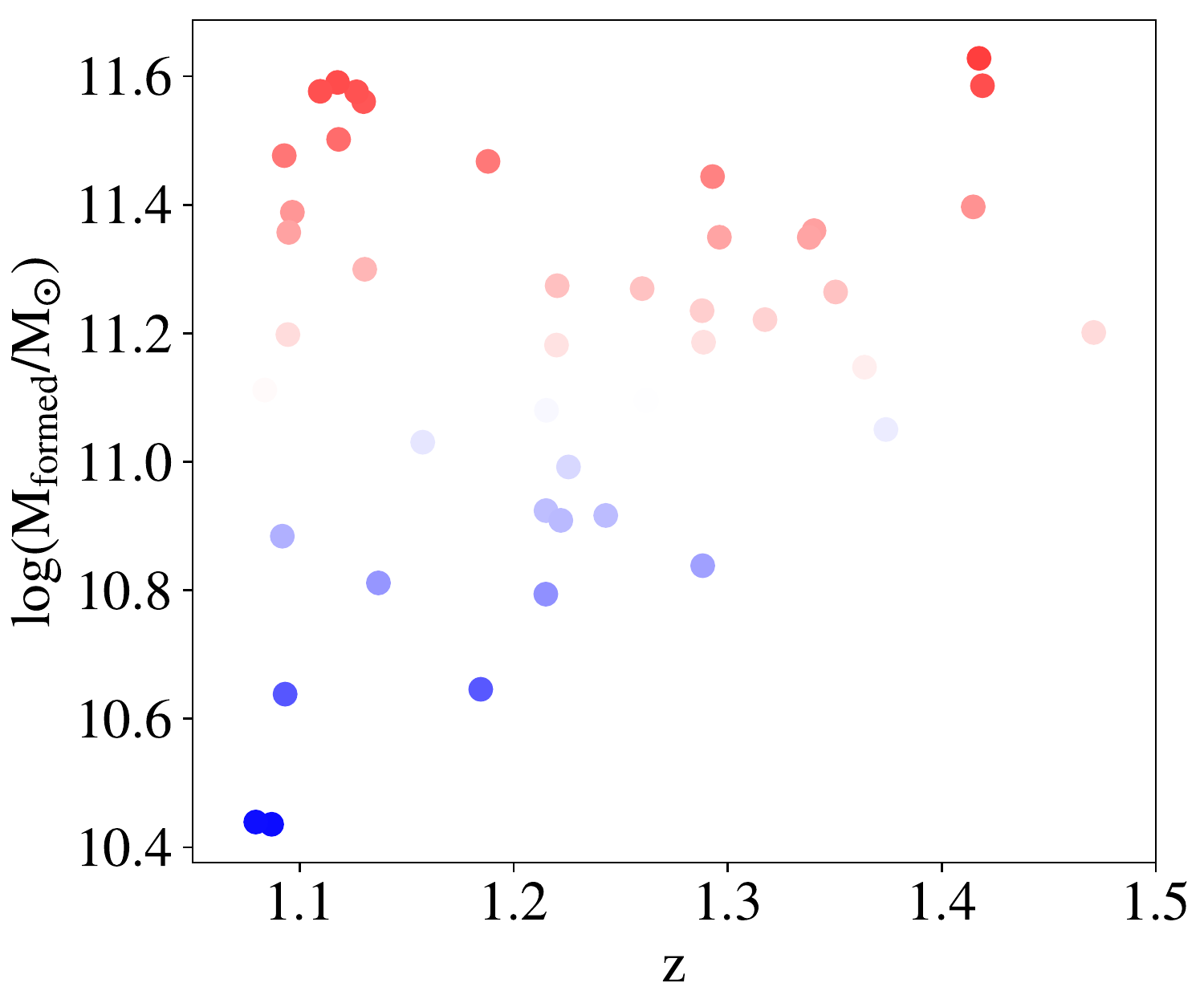}
         \caption{}
         \label{fig:mass_z}
     \end{subfigure}
     \hfill
     \begin{subfigure}[b]{0.49\textwidth}
         \centering
         \includegraphics[width=0.9\textwidth]{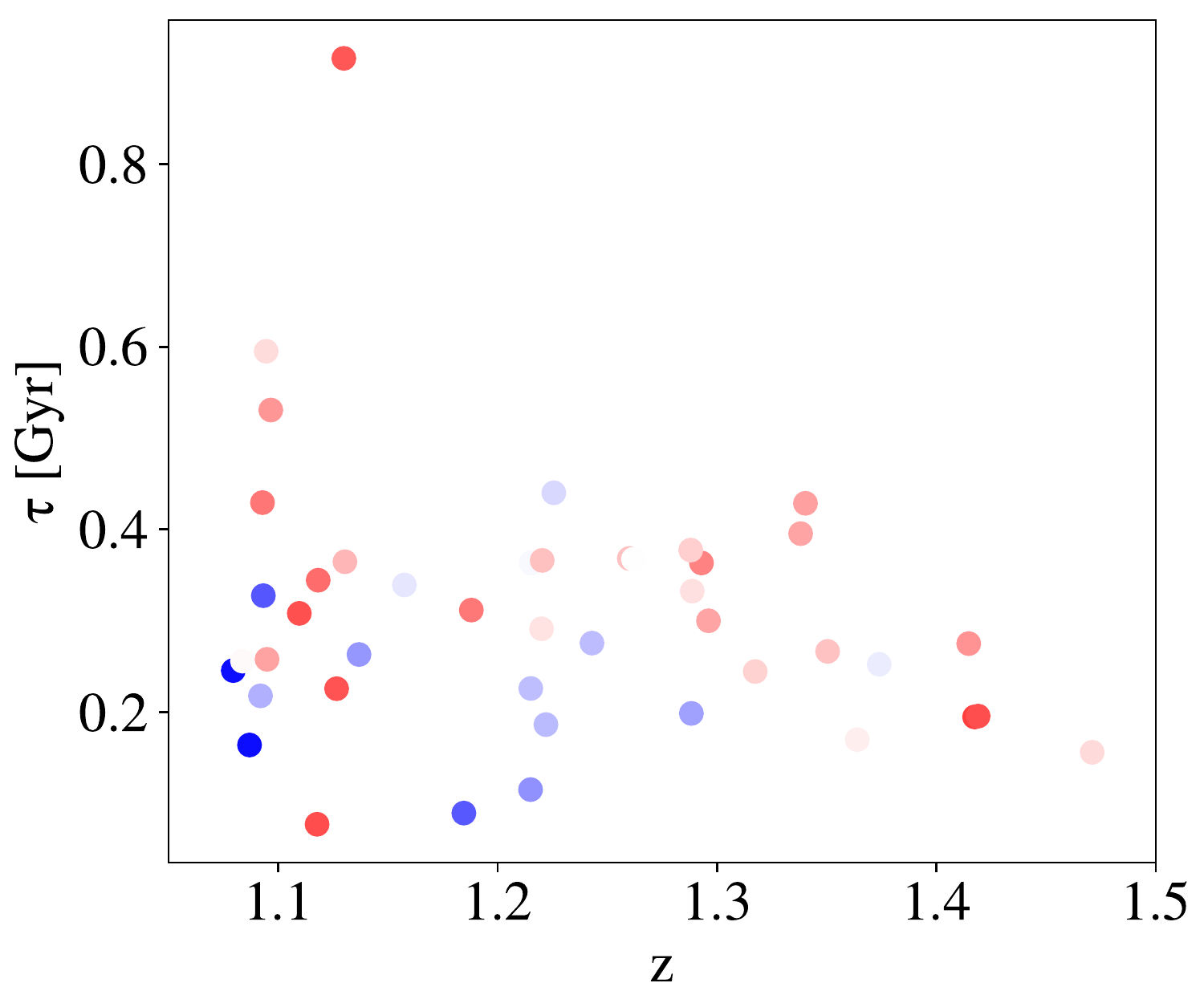}
         \caption{}
         \label{fig:tau_z}
     \end{subfigure}
     \hfill
     \begin{subfigure}[b]{0.5\textwidth}
         \centering
         \includegraphics[width=0.9\textwidth]{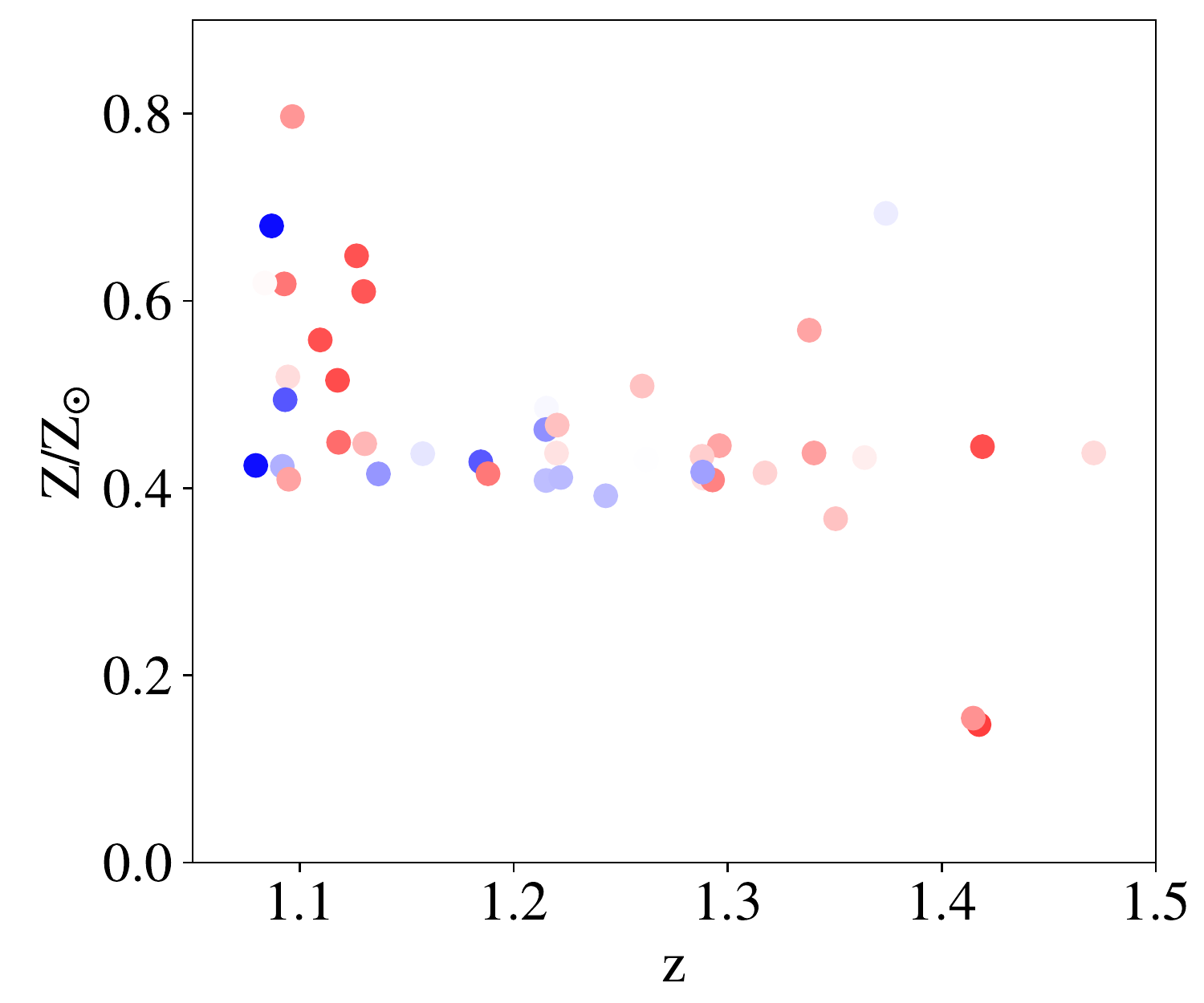}
         \caption{}
         \label{fig:met_z}
     \end{subfigure}
     \hfill
        \caption{Trends with redshift of age (a), stellar mass (b), tau parameter (c) and metallicity (d), colour-coded by stellar mass. In the age-redshift the lines in the background represent the theoretical trends in a flat$\Lambda$CDM cosmology with $\Omega_{m,0}$ = 0.3, $\Omega_{\Lambda,0}$ = 0.7, $\mathrm{H_0}$ = 70 km/s/Mpc with different formation redshifts.}
        \label{fig:3var_z}
\end{figure*}

By adopting an archaeological look-back approach, we can study how our measured quantities vary as a function of redshift. In particular, given the cosmological purpose of this work, we are interested in the age-redshift relation, that will be discussed in detail in Sect. \ref{sec:4.1OptimisingCC}. In Fig. \ref{fig:age_z} we show the age-redshift for our 44 CCs coloured by stellar mass. From this trend, we can draw three main conclusions. First, for most of the galaxies (95\% of CCs), ages are lower than the age of the Universe as expected in a standard cosmology (flat$\Lambda$CDM with $\Omega_{m,0}$ = 0.3, $\Omega_{\Lambda,0}$ = 0.7, $H_0$ = 70 km/s/Mpc), even without having imposed a cosmological prior, since they could formally vary between 0 and 20 Gyr across the entire redshift range. This is \textit{per se} a significant result, since it demonstrates that with enough spectral coverage, S/N and photometric data we can constrain reliably this parameter without any additional cosmological assumption. Second, the ages show a decreasing trend with redshift in agreement with the cosmological expectation, in line with what was already observed for the $\mathrm{D_n4000}$. The third important observation is the existence of a mass-downsizing trend, with more massive galaxies being older than the less massive ones at fixed redshift. In particular, objects with $\mathrm{\massf}\leq 11.1$ show formation redshifts in the range $\mathrm{1.5 \lesssim z_{form} \lesssim 4}$ while for galaxies with $\mathrm{\massf}>11.1$ it spans the range $\mathrm{2 \lesssim z_{form} \lesssim 7}$.

In Fig. \ref{fig:3var_z} we show the trends with redshift for other relevant physical parameters estimated in the fit (stellar mass, metallicity and $\tau$ parameter) coloured accordingly to the stellar mass. We can notice that for the more massive sample (in red) there are no strong trends with redshift for any of the considered quantities, except the ages, while the lower mass sample (in blue) shows a mild increasing trend with redshift for the stellar mass. This is a well-known effect due to the observational luminosity threshold, which allows us to see only the intrinsically brightest objects when we observe the distant Universe. This effect can be noted in particular for the stellar mass since it correlates with the galaxy luminosity. Given this, it's clear that to obtain a homogeneous sample as a function of redshift for the cosmological analyses, it will be necessary to perform a cut in stellar mass, as will be discussed in Sect. \ref{sec:4.1OptimisingCC}.

Another important observation in this context can be derived for the $\tau$ parameter, which is related to the SFH length, and its trend with redshift in Fig. \ref{fig:tau_z}. Both for the higher and the lower mass galaxies it is stable around 0.3 Gyr, but the high-mass sample seems to have a longer period of star formation, in contrast with what is expected in a mass-downsizing paradigm, in which the more massive a galaxy is, the older it is expected to be and also the shorter its SFH. We trace the origin for this inversion to be the existent degeneracy among $\tau$ and age, occurring as a direct correlation between these two parameters. This effect is well known in literature \citep{Gavazzi2002} when adopting a DED SFH, since age and $\tau$ have very similar effects on the spectral shape and it is difficult to disentangle their contribution. 
The impact of this effect, however, is negligible on our results, since the range of retrieved $\tau$ is extremely small and compatible with a very short SFH, and, most importantly, because our cosmological analysis is made on sub-samples with constant mass where the $\tau$ appears very stable as a function of redshift. 

In this context, it is even more interesting to analyse the impact on the results of adopting a DPL SFH, in which the decoupling of the rising and falling slope allows more freedom to the SFH shape.\\
The assumed SFH has been identified in recent literature as a potentially significant source of systematic effects in the estimate of galaxies' physical parameters. For this reason, we extensively test the dependence of our measurements on the assumed SFH functional form. In particular, we repeat our analysis considering the DPL SFH (config. 1) and assess the robustness of our results. We provide in Appendix \ref{appendixB} the detailed discussion of our findings. In summary, we obtain that the age estimates are very robust, with minimal variations occurring when changing the SFH. In particular, setting a lower limit to the rising slope of the DPL SFH ($\beta>10$) to exclude non-physical solutions, we find a median percent difference in age estimates below 2\%. In terms of velocity dispersion and dust reddening, the discrepancy is even smaller, reaching a median percent difference lower than 1\%, while the gap in stellar mass is of 0.001 dex. Moreover, the resulting SFH shapes are almost identical when adopting a DED or a DPL, despite the different functional forms. In the following sections, we will use the results of config. 1 to assess the systematic effect introduced by assuming a different SFH in estimating the cosmological parameters.

\section{COSMOLOGICAL ANALYSIS}
\label{sec:4Cosmo}
In this section we first present how we build the median age-redshift relation, how it is fitted with a flat $\Lambda$CDM cosmology to derive constraints on the Hubble constant $\mathrm{H_0}$, and finally the cosmological constraints derived from the application of the CC method.

\subsection{Optimising the selection of cosmic chronometers}
\label{sec:4.1OptimisingCC}
In Sect. \ref{sec:3.3PhysProp} the discussion of the physical properties of the 44 selected galaxies has provided additional support to the fact that these objects meet the conditions for being CCs, given their high masses ($\mathrm{\langle \massf \rangle=11.21 \pm 0.05}$) and their very short periods of star formation $\mathrm{\langle \tau \rangle = 0.28 \pm 0.02}$ Gyr. At this point, an important aspect to be considered in applying the method is to maximise the synchronisation of the sample formation time. In a mass-downsizing scenario high-mass galaxies \citep[the cut $\mathrm{\mass > 10.6}$ is often applied,][]{Moresco2022} are the first to form \citep[z>2-3,][]{Citro2017,Carnall2018,Carnall2019} in a short burst of star formation \citep[t<0.3 Gyr,][]{Thomas2010,Carnall2018}, so are the best to provide a sample of synchronised chronometers. Anyway, it is important not only to select massive galaxies but also that their properties are consistent in redshift, in order to maximise the homogeneity of the sample at different cosmic times and avoid biases in the cosmological analysis. In the previous section, commenting Fig. \ref{fig:3var_z}, we noted that the sample shows homogeneous metallicity and $\tau$ throughout redshift, while the stellar mass for the low-mass sample, in blue in Fig. \ref{fig:mass_z}, shows an increasing trend with redshift. Then, aiming to uniform the sample mass, we apply a further cut $\mathrm{\massf}$<10.8 (discarding 5 galaxies), a threshold that not only makes our CCs homogeneous in mass, but also in redshift of formation, removing all the galaxies with $\mathrm{z_{form}}$<2 in Fig. \ref{fig:age_z}.\\
The sample of CCs that we adopt for the cosmological analysis, then, counts 39 galaxies.

\begin{figure}
        \centering
        \includegraphics[width=0.4\textwidth]{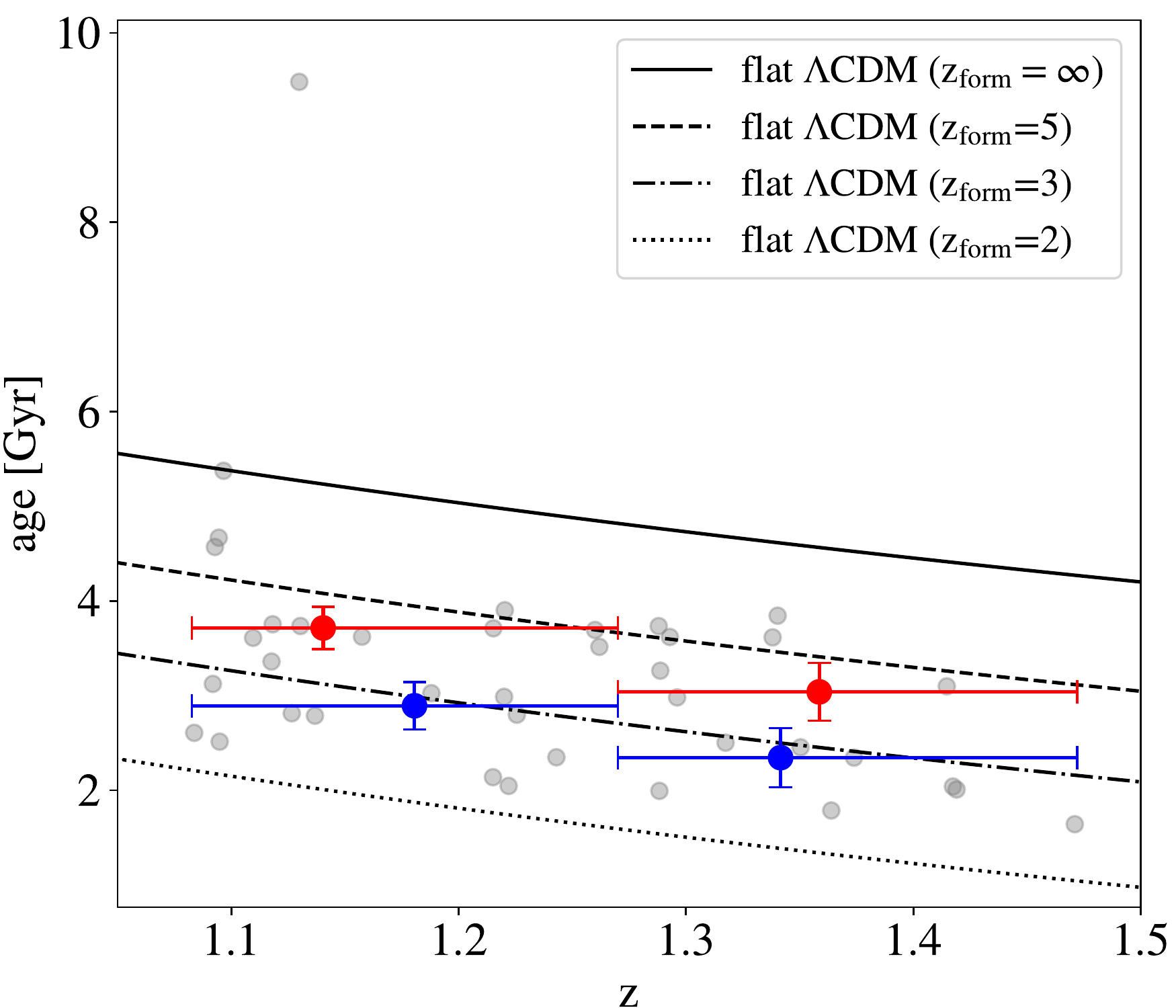}
    \caption{Median age-redshift trend for the 39 CCs sample, obtained with binning A. Red and blue dots represent median values respectively for the high-mass ($\mathrm{\massf>11.26}$) and the low-mass ($\mathrm{\massf\leq11.26}$) sample. In the background, grey dots are the 39 single measurements while lines show, for illustrative purposes only, the theoretical trends with different redshifts of formation, as given by a flat $\Lambda$CDM model with $\mathrm{\lcdm}$.}
    \label{fig:4age_zA}
\end{figure}

With this accurately selected sample, we can now build the median age-redshift relation, which is more robust in tracing the ageing trend since it allows us to increase the S/N of our measurements. This consists in dividing the sample into redshift bins and eventually into mass bins, then averaging ages and redshifts in each subgroup. To each median age, we associate an error computed as $\mathrm{MAD/\sqrt{N}}$. We test different types of binning:
\begin{itemize}
    \item equally spaced or equally populated in redshift;
    \item two or four redshift bins;
    \item subdividing (or not) the sample into mass bins according to the median value of the $\mathrm{\massf}$ distribution ($\mathrm{\langle \massf \rangle = 11.26}$).
\end{itemize}
When dividing the sample into mass sub-samples, we decide to split them in redshift bins after the mass separation, to obtain more homogeneous sub-samples in mass, and hence time of formation.
By combining these three options, eight different binning types can be obtained, but not all of them are effective in tracing the age-redshift trend. For example, using the separation in mass and four redshift intervals at the same time, produces a total of eight sub-samples with $\leq$5 galaxies each, ending with median values that are very sensitive to fluctuations in each bin. On the contrary, adopting two redshift bins and no mass separation would lead to more stable median values, but having just a pair of points is not effective in constraining the age-redshift trend. Additionally, aiming for the maximum synchronicity of the population that we average on, we find it better to always adopt the mass separation, which guarantees a higher homogeneity of the sample in each bin.\\
After these considerations, we can conclude that the best binning types for our sample are two, given by two equally spaced or equally populated redshift bins, divided in mass. We refer to the equally spaced one as \textit{binning A}, which will be used as a benchmark, while the equally populated one is referred to as \textit{binning B} and used as a comparison. This choice is made because the first one guarantees a more homogeneous sampling of the ageing trend also in terms of redshift. The median age-redshift trend for binning A is shown in Fig. \ref{fig:4age_zA} and the relative median values and errors are reported in Tab. \ref{tab:4binningA}.

\begin{table}
    \centering
    \caption{Median ages and properties of the selected CC sample considering the binning A as described in the text. For each sub-sample are reported: mean values of redshift, median ages, median stellar mass and the number of galaxies in each bin.}
    \begin{tabular}{c c c c c}
        \hline\hline
        Binning A   & z & age [Gyr] & $\mathrm{\massf}$ & N\\ \hline
        \multirow{2}{*}{high mass} & 1.14 & 3.7 ± 0.2 & 11.47 ± 0.04 & 12 \\
                    & 1.36 & 3.0 ± 0.3 & 11.38 ± 0.02 & 8  \\ 
        \multirow{2}{*}{low mass} & 1.18 & 2.9 ± 0.2 & 11.01 ± 0.04 & 12 \\
                    & 1.34 & 2.3 ± 0.3 & 11.19 ± 0.02 & 7  \\
        \hline
    \end{tabular}
    \label{tab:4binningA}
\end{table}

\subsection{Analysing the age-redshift relation}
\label{sec:4.2AnalysingAgeZ}
As a first test, we directly study the obtained median age-redshift relation. In particular, we fit the median age-redshift with a flat $\Lambda$CDM model, where the free parameters are the Hubble constant $\mathrm{H_0}$ and the adimensional matter density parameter ($\mathrm{\Omega_{M,0}}$). In this framework, the Hubble parameter \textit{H(z)} can be expressed as:
\begin{equation}
    \mathrm{H(z) = H_0 \sqrt{1-\Omega_{M,0} + \Omega_{M,0}(1+z)^3}}.
\end{equation}
In addition, considering a Friedmann-Lemaitre-Robertson-Walker (FLRW) metric, the age of the Universe at a given redshift \textit{t(z)} is linked to the Hubble parameter through the equation:
\begin{equation}
    \mathrm{t(z) = \int_0^z \frac{dz'}{H(z')(1+z')}}.
    \label{eq:4t(z)}
\end{equation}
Before fitting this age-redshift relation to our median data we need to pay attention to the fact that \textit{t(z)} here refers to the age of the Universe, while we obtained ages for a sample of CCs. This means that the two trends are separated by an offset, due to the delay between the Big Bang and the formation of the first galaxies. We parameterise this offset as $t_0$, representing the age of the Universe at which our CCs formed. Assuming that the objects in the sample are coeval, $t_0$ can be considered constant in redshift. Its value, instead, will be left free to vary. Introducing $t_0$, Eq. \ref{eq:4t(z)} can be modified to suit the CCs age-redshift trend as follows:
\begin{equation}
    \mathrm{t_{cc}(z) = \int_0^z \frac{1}{1+z'}\frac{dz'}{H_0 \sqrt{1-\Omega_{M,0} +\Omega_{M,0}(1+z')^3}}-t_0}.
    \label{eq:4tCC(z)}
\end{equation}
Constraining these three parameters at the same time is a non-trivial process, mainly due to their degeneracies: for example, a higher $t_0$ results in a lower age of the Universe, but this happens with a higher $H_0$ or a larger $\Omega_{0,M}$ too. As it is also shown in \citet{Borghi2022}, even if these degeneracies are non-negligible, the three parameters affect differently the age-redshift slope. So, if age errors are small enough and if appropriate priors on the parameters are adopted, we could mitigate, at least partially, these degeneracies.

We perform the fit of the age-redshift relation obtained with binning A (shown in Fig. \ref{fig:4age_zA} and described in Tab. \ref{tab:4binningA}) with the theoretical trend in Eq. \ref{eq:4tCC(z)} adopting a Markov Chain Monte Carlo (MCMC) technique implemented through the affine-invariant \texttt{emcee} sampler \citep{ForemanMackey2013} and considering a Gaussian likelihood function. Uniform, uninformative priors are assumed for $H_0$ and $t_0$: $H_0 \sim \mathcal{U}(25,125)$, $t_0 \sim \mathcal{U}(0.5,5)$. For $\Omega_{M,0}$, instead, a Gaussian prior is adopted: $\Omega_{M,0} \sim \mathcal{G}(0.3,0.02)$, required to keep degeneracies under control. Its value and uncertainty refer to the ones used in \citet{Jimenez2019}, resulting from the combination of different measurements of $\Omega_{M,0}$, all independent of the Cosmic Microwave Background (CMB).

If the mass-downsizing scenario is valid, lower-mass galaxies should have formed later than higher-mass ones. Since the median age-redshift relation is divided into a high-mass and a low-mass sample, to keep this effect into account, we introduced a parameter $\Delta t$ that measures the offset in formation time between the two populations. Differently from $t_0$, $\Delta t$ is not considered a free parameter but its value is assumed constant, computed as the average age separation between these two sub-samples. For binning A we find $\Delta t \sim 0.8$ Gyr. To jointly fit all data, therefore, we consider this shift of the lower mass sample in our analysis.

\subsubsection{Results}
\begin{figure}
    \centering
    \includegraphics[width=0.4\textwidth]{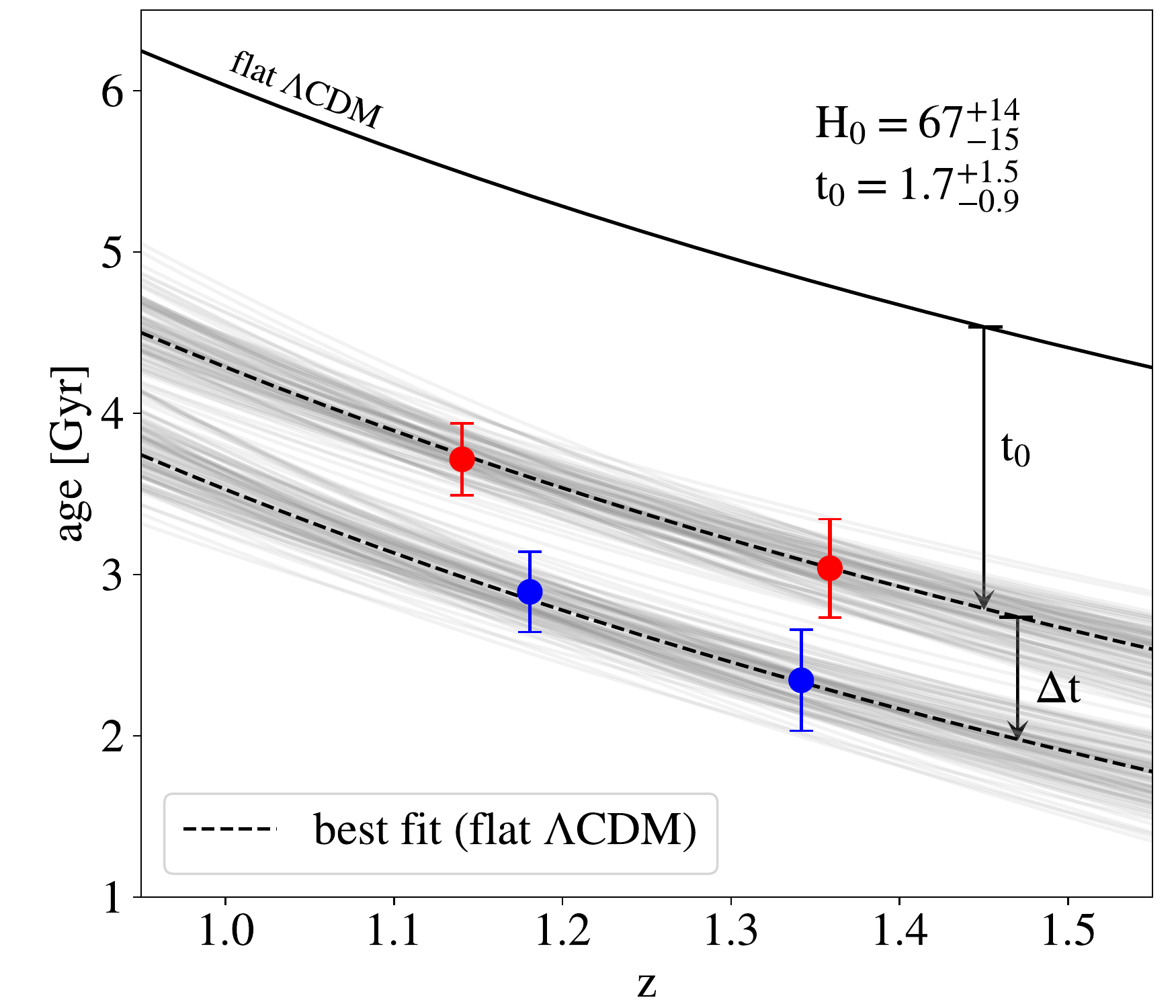}
    \caption{Fit to the median age-redshift trend obtained with binning A. Red and blue dots are median values respectively for the high-mass ($\mathrm{\massf>11.26}$) and the low-mass ($\mathrm{\massf\leq11.26}$) sample. In grey are shown flat $\Lambda$CDM trends with ($\mathrm{H_0,\Omega_{M,0}},t_0$) values randomly extracted from the posterior distribution, comprised between the $\mathrm{16^{th}\: and\: the\: 84^{th}}$ percentiles. Dashed black lines represent the best fit.}
    \label{fig:4age_z_fit}
\end{figure}

Performing the fit we obtain $H_0\: \mathrm{= 67^{+14}_{-15}\: km\:s^{-1}\:Mpc^{-1}}$ and $t_0\: \mathrm{= 1.7^{+1.5}_{-0.9}\: Gyr}$. Values and errors are respectively medians and 1$\sigma$ values of the posterior distribution. Results are shown in Fig. \ref{fig:4age_z_fit}, where grey curves are drawn from the posterior distribution of parameters, between the $\mathrm{16^{th}\: and\: the\: 84^{th}}$ percentiles.

Comparing these results with the ones in \citet{Riess2022} and \citet{PlanckCollaboration2020} we can say that our errors are not conclusive enough to prefer one or the other, and our measurements agree both with the late- and the early-Universe estimations of $H_0$. We note here that the large error on $\mathrm{H_0}$ is mostly due to the very low statistic of CCs available in this analysis.\\
However, these results are still promising in view of upcoming large surveys, like Euclid \citep{Laureijs2011}, where both the redshift coverage and the much lower statistical errors, granted by the much higher statistics, could significantly increase the precision of the cosmological parameters.

\subsection{Cosmological constraints with the cosmic chronometers approach}
\label{sec:4.3CCapproach}
Starting from the median age-redshift relation obtained in Sect. \ref{sec:4.1OptimisingCC} we can now apply the cosmic chronometers method. As already discussed in Sect. \ref{sec:1Intro}, the Hubble parameter $H(z)$ can be estimated by evaluating the differential age evolution over a small redshift bin dz of a sample of cosmic chronometers through the equation:
\begin{equation}
    \mathrm{H(z) = -\frac{1}{1+z} \frac{dz}{dt}}
    \label{eq:4H(z)}
\end{equation}
which means that we can estimate the Hubble parameter at the redshift of the CCs sample by measuring the first derivative of the age-redshift relation. This requires:
\begin{itemize}
    \item an accurately synchronised sample of CCs;
    \item a suitably large redshift interval, such that the differential in age (dt) is larger than its uncertainty;
    \item small age errors, to maximise the accuracy on the measurement of dz/dt.
\end{itemize}
The selection process presented in Sect. \ref{sec:2.2SelectionCosmicChronometers} and the optimisation of the CCs sample in Sect. \ref{sec:4.1OptimisingCC} showed that the 39 selected CCs have the characteristics to satisfy the first requirement. Moreover, the chosen binning type (binning A, described in Sect. \ref{sec:4.1OptimisingCC}) is optimal for meeting these requisites, representing a good balance between the need for sampling in redshift and the need for statistics in each bin. At the same time, adopting the mass separation, it is able to maximise the synchronicity of each sub-sample given the strong correlation among stellar mass, age and redshift of formation expected in a mass-downsizing scenario \citep{Thomas2010}.\\
Starting from the median age-redshift relation obtained with binning A, shown in Fig. \ref{fig:4age_zA} and reported in Tab. \ref{tab:4binningA}, the Hubble parameter $H(z)$ is computed in two steps: first, Eq. \ref{eq:4H(z)} is applied separately on the two high- and low-mass data points, where $z$ is computed as the mean redshift of the two points, obtaining two $H(z)$ measurements and relative errors; then we compute the average of these two values, weighted on the associated error. We can perform this average because, while the median age in the two mass sub-samples is clearly offset due to a different time of formation, and hence needs to be analysed separately, the underlying cosmology has to be the same, and therefore the Hubble parameter estimates can be averaged to increase the accuracy of the measurement. In this way, we obtain an estimate for the Hubble parameter at z$\simeq$1.26:
\begin{equation*}
    \mathrm{H(z\simeq1.26) = 135 \pm 60\: (stat) \quad [km\:s^{-1}\:Mpc^{-1}]}
\end{equation*}
where the associated error is given here by the only contribution of the statistical uncertainty, resulting from the propagation of the error on median ages, which scales with $\mathrm{\sqrt{N}}$, the number of elements in each bin. In the next section, we are going to analyse the impact on the result of two effects, the binning choice and the SFH choice, aiming to include a systematic component in the uncertainty.

\subsubsection{Study of the systematic effects on H(z)}
In Sect. \ref{sec:3.1FSFwithBagpipes} we discussed the different types of fit configurations that have been tested on our CCs sample. We paid particular attention to config. 1, which differs from the baseline in having a DPL SFH instead of a DED SFH, to which Appendix \ref{appendixB} is dedicated. There, we found a strong agreement between baseline and config. 1 results with a mean percent difference in ages smaller than 2\%. Here we are interested in understanding how this slight discrepancy in age propagates to the estimation of the Hubble parameter. We should recall here that when using a DPL SFH, \texttt{BAGPIPES} provides only mass-weighted ages, as defined in Eq. \ref{eq:3age_MW}, while baseline results are built with standard ages, from the definition in Eq. \ref{eq:3age}. As already discussed, this difference does not introduce a change in the slope of the ageing trend, because we verify that the offset between the two definitions is constant in redshift.\\
After performing the fit with configuration 1, we discard the five objects identified in Sect. \ref{sec:4.1OptimisingCC} to have $\mathrm{\massf <10.8}$ and then we clear up the sample from bad fits, ending up with a sample of 36 CCs. With these, we obtain a new median age-redshift applying binning A, and then implement the CC method repeating the process explained in Sect. \ref{sec:4.3CCapproach}. The Hubble parameter measurement, in this case, is $H \mathrm{(z\simeq1.25) = 156 \pm 51}$.

To include in the total error budget also the effect of the assumed binning, we repeat this process also for binning B, both with baseline and configuration 1 results. The four measurements obtained are reported in Tab. \ref{tab:4sys}. To estimate the impact of choosing a different SFH, we compute the average difference between baseline and configuration 1 measurement of $H(z)$ for equivalent binnings, which results in a contribution to the error budget of $\mathrm{\Delta H_{SFH} = 27\: km\:s^{-1}\:Mpc^{-1}}$. We note that this large uncertainty is dominated by the $H(z)$ value obtained with config. 1 and binning B, the less accurate of the four estimates that we derive, as shown in Tab. \ref{tab:4sys}. 
This effect is due to the fact that in this configuration the number of correctly estimated ages is smaller than in the baseline configuration, and therefore, being the statistics lower, it is more subject to fluctuations when varying the binning. In particular, we notice that by choosing the equally populated binning we end up whit an uneven redshift sampling that increases the fluctuation in the average ages, resulting in a significantly larger uncertainty on H(z). If we exclude this result, $\mathrm{\Delta H_{SFH}}$ would be cut down to $\mathrm{ = 15\: km\:s^{-1}\:Mpc^{-1}}$. Lastly, we include in the error also the discrepancy in $H(z)$ between baseline binning A and binning B, $\mathrm{\Delta H_{bin} = 2.4\: km\:s^{-1}\:Mpc^{-1}}$. Adding these two contributions we obtain:
\begin{align*}
    &\mathrm{H(z\simeq1.26) = 135 \pm 60\: (stat)} \\ 
    &\mathrm{\quad\quad\quad\quad\quad\quad \pm 27\: (sys) \pm 2.4\:(bin) \quad [km\:s^{-1}\:Mpc^{-1}]}
\end{align*}
and finally, summing the errors in quadrature:
\begin{equation*}
    \mathrm{H(z\simeq1.26) = 135 \pm 65 \quad [km\:s^{-1}\:Mpc^{-1}]}.
\end{equation*}

\begin{table}
    \centering
    \caption{Measurements of H(z) obtained applying the CC method to median age-redshift trends obtained with baseline and configuration 1 fits, both using binning A and binning B.}
    \begin{tabular}{c c c c}
    \hline\hline
        & \textbf{binning} & $\mathbf{\langle z \rangle}$ & $\mathbf{H(z) [km\:s^-1\:Mpc^{-1}]}$ \\ \midrule
        \multirow{2}{*}{baseline}  & A & 1.26 & 135 $\pm$ 60 \\
                                   & B & 1.24 & 132 $\pm$ 82  \\ 
        \multirow{2}{*}{config. 1} & A & 1.25 & 120 $\pm$ 34 \\
                                   & B & 1.27 & 171 $\pm$ 109 \\ \bottomrule  
    \end{tabular}
    \label{tab:4sys}
\end{table}

\begin{figure*}
    \centering
    \includegraphics[width=0.95\textwidth]{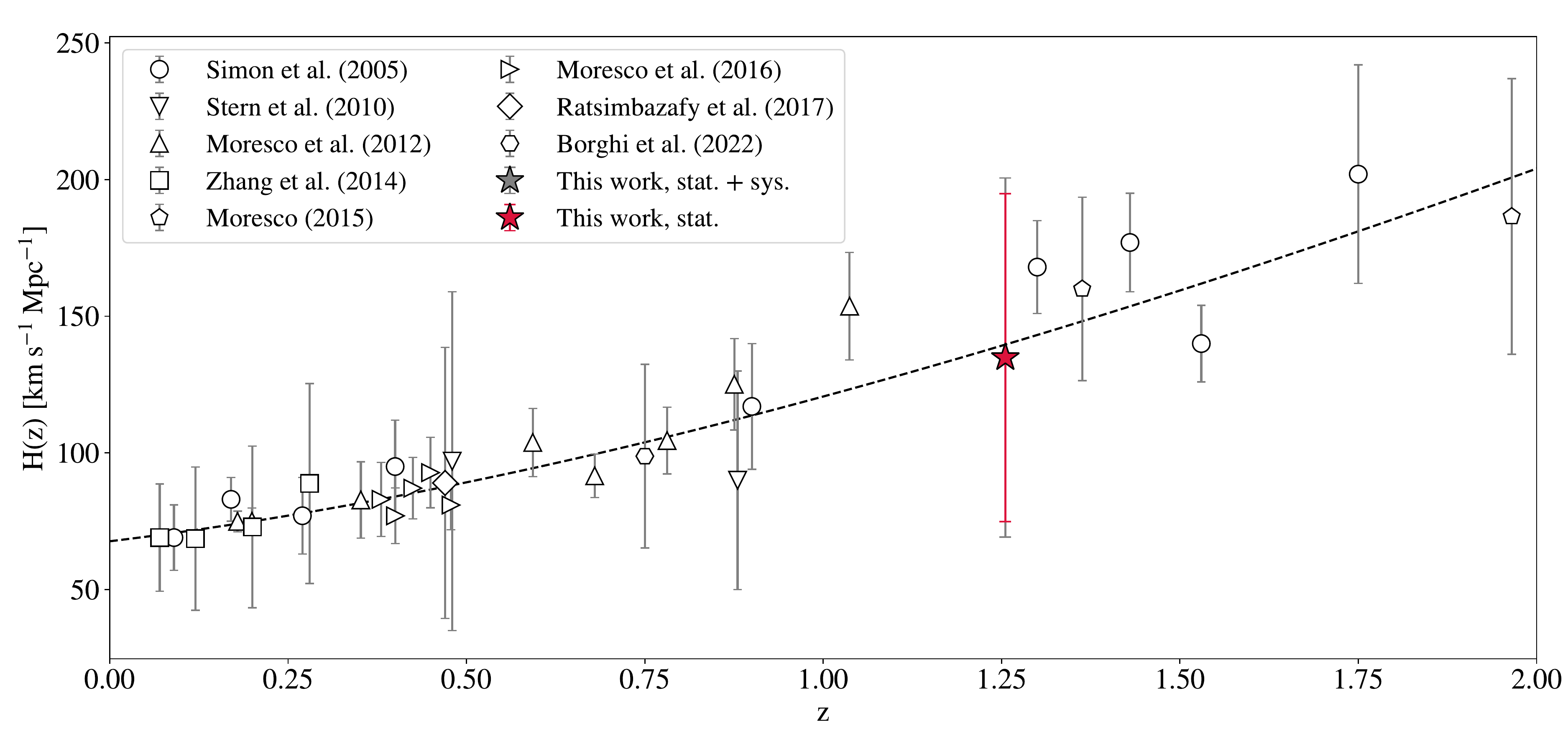}
    \caption{H(z) measurement obtained in this work in comparison with all the H(z) estimations obtained up to now with the cosmic chronometers method. The dashed line represents the theoretical trend of a f$\Lambda$CDM model as in \citet{PlanckCollaboration2020}, as a pure illustrative reference.}
    \label{fig:4H(z)_CC}
\end{figure*}

\noindent
This measurement, which represents the main result of this work, is also shown in Fig. \ref{fig:4H(z)_CC}, where all the $H(z)$ measurements obtained with the CC method are reported as a function of redshift. 
It is clear that the statistical uncertainty, here, is dominating the error budget of the final result, as a direct consequence of the low number of CCs in the sample. Recalling that the statistical error on $H(z)$ depends on the uncertainty of median ages, computed as $\mathrm{MAD/\sqrt{N}}$, to cut down its contribution we would need larger statistics. For comparison, in \citet{Jiao2022} an estimate of the Hubble parameter $\mathrm{H(z\simeq0.80) = 113.1 \pm 15.2 (stat)}$ was obtained with a sample of 350 CCs from the LEGA-C survey \citep{Straatman2018}, reaching a statistical error of 13\%. Indeed, starting from the statistical error obtained in this work of 44\% with a sample of 39 objects and assuming that this scales with $\mathrm{\sqrt{N}}$, the expected statistical error for a sample of 350 galaxies is around 15\%, very similar to the 13\% actually obtained.  This gives an optimistic prospect on the achievable results with future, much larger, surveys. The ESA space mission Euclid \citep{Laureijs2011}, for example, is expected to observe up to a few thousand CC candidates in the redshift range 1.5<z<2 increasing the current statistics by two orders of magnitude. In the context of $H(z)$ estimate with the CC method, this would mean being able to obtain measurements of the Hubble parameter up to z=2 with a statistical error of the order of 6\%, increasing significantly the precision of the estimates at this redshift which stands now between 10\% and 25\%. In this way, we could constrain with higher accuracy the $H(z)$ trend and, as a consequence, the expansion history of the Universe.

\section{Conclusions}
\label{sec:5CONCLUSIONS}
In this work, we select and analyse a sample of massive and passively evolving galaxies from data release 4 of the VANDELS spectroscopic survey \citep{McLure2018}, with the purpose of estimating the Hubble parameter $H(z)$ with the cosmic chronometers method. We adopt a full-spectral-fitting technique to estimate the physical properties of the sample, to benefit from all the spectral and photometric information available in VANDELS, using the code \texttt{BAGPIPES} \citep{Carnall2018}, already adopted and optimised within the survey. Differently from the studies already carried out in VANDELS, in this work we adopt a modified version of the code introduced in Jiao et al. (2022), where a non-cosmological prior is assumed on the age of the galaxy population and at all redshift is free to vary between 0 and 20 Gyr. This means that the resulting ages are not constrained by a chosen cosmology, but depend only on the adopted stellar population synthesis models and on the component used in the fit to reproduce spectra and photometry.\\
Our main results are summarised as follows:
\begin{enumerate}
    \item We select a sample of 49 purely passive galaxies in the redshift range 1<z<1.5 adopting multiple and complementary selection criteria: a photometric criterion (UVJ), a cut on the [OII] emission line, a cut on the H/K ratio, a visual inspection and also a redshift cut. This last one is needed because of an anomaly found in the D4000 trend, for which galaxies at z<1.07 are discarded to avoid biases in the subsequent analysis. The selected galaxies show a red continuum, no emission lines linked to stellar activity, and no H/K inversion. They turn out to have high stellar masses and low specific star formation rates, with median values of $\mathrm{\langle \mass \rangle = 10.88 \pm 0.05}$ and $\mathrm{\langle \ssfr \rangle = -12.2 \pm 0.2}$;
    \smallskip
    \item Studying the evolution of age-related spectral features, mainly D4000 and $\mathrm{D_{n}4000}$, we observe a clear decreasing trend with redshift. Since these features are proven to correlate with the galaxy age, this gives a first, purely observational, evidence that the population under analysis is ageing with cosmic time. We also observe that, at fixed redshift, more massive galaxies show a higher D4000 with respect to the lower mass ones, supporting a mass-downsizing scenario;
    \smallskip
    \item Performing full-spectral-fitting, we obtain age estimates smaller than the age of the Universe in a standard flat $\Lambda$CDM model ($\mathrm{\lcdm}$) for 95\% of the sample, even if they could formally vary between 0 and 20 Gyr. This proves the robustness of our estimates, where the good quality of VANDELS data allows us to determine correct ages even without imposing the standard cosmological prior on them. Ages also decrease with redshift in agreement with this cosmological model and, at fixed redshift, more massive galaxies result older than lower mass ones, confirming the mass-downsizing trend.\\
    The sample is characterised by high masses, sub-solar metallicities, low dust extinction, and a short period of star formation, with median values of mass, metallicity, dust reddening and $\tau$ parameter equal to: $\mathrm{\langle \massf \rangle = 11.21 \pm 0.05}$, $\mathrm{\langle Z/Z_{\odot} \rangle = 0.44 \pm 0.01}$, $\mathrm{\langle A_{V,dust} \rangle = 0.43 \pm 0.02}$ mag and $\mathrm{\langle \tau \rangle = 0.28 \pm 0.02}$ Gyr. Metallicity values are compatible with those obtained in \citet{Carnall2022} on a sample of VANDELS passive galaxies.
    \smallskip
    \item Comparing results obtained by fitting with a delayed SFH or a double-power law SFH we find that, despite the different functional forms, the two turn out to be substantially identical if the involved parameters are constrained in a physically reasonable range. In particular, by setting a lower limit to the rising slope of the double-power law SFH, $\mathrm{\beta >10}$, the median percent difference in age and metallicity estimates is below 2\%, for dust reddening and velocity dispersion is less than 1\%, while for stellar mass is of 0.001 dex.
    \smallskip
    \item We further clean the 49 galaxies sample by removing bad fits and applying a mass cut ($\mathrm{\massf\leq10.8}$) to homogenise it, ending with a sample of 39 cosmic chronometers. We build a median age-redshift relation by dividing the sample into two mass bins and two redshift bins. By fitting this median age-redshift relation with a flat $\Lambda$CDM model, we obtain an estimate for the Hubble constant $H_0$ = $\mathrm{67^{+14}_{-15} km\: s^{-1}\: Mpc^{-1}}$ and for the formation time of high-mass objects $\mathrm{t_0 = 1.7^{+1.5}_{-0.9}}$ Gyr. In doing this, we need to set a Gaussian prior on $\mathrm{\Omega_{M,0} = 0.30 \pm 0.02}$ in order to keep degeneracies under control.
    \smallskip
    \item Finally, we obtain a new, cosmology-independent, direct measurement of the Hubble parameter at z$\sim$1.26 equal to $\mathrm{H(z) = 135}$ $\mathrm{\pm 65\: km\:s^{-1}\:Mpc^{-1}}$ applying the cosmic chronometers method. Errors include both statistical and systematic uncertainties, with the first dominating the error budget. In the systematic component, we include the effect on the $H(z)$ estimate by using a more complex SFH and the effect of changing the binning while computing the median age-redshift relation.
\end{enumerate}

Concluding, this work provides additional evidence supporting the robustness of the CC method up to z=1.5 while proving the effectiveness, even at this redshift, of adopting a full-spectral-fitting approach to extract ages and physical parameters of the galaxy population.\\
At the same time, the positive results obtained in terms of $H(z)$ and $H_0$, despite the poor statistics, are very promising in view of upcoming large spectroscopic surveys like Euclid \citep{Laureijs2011}. Forecasting a number of cosmic chronometer candidates of two orders of magnitude greater than the ones in this work, we could expect to bring down the statistical error to 6\% up to $\mathrm{z\sim2}$.

\begin{acknowledgements}
    MM and ACim acknowledge the grants ASI n.I/023/12/0 and ASI n.2018-23-HH.0. ACim acknowledges the support from grant PRIN MIUR 2017 - 20173ML3WW\_001. MM acknowledges support from MIUR, PRIN 2017 (grant 20179ZF5KS). ET acknowledges the support from COST Action CA21136 – “Addressing observational tensions in cosmology with systematics and fundamental physics (CosmoVerse)”, supported by COST (European Cooperation in Science and Technology). ACCar thanks the Leverhulme Trust for their support via a Leverhulme Early Career Fellowship.
    
\end{acknowledgements}

%
%
\bibliographystyle{aa}
\bibliography{references}

\appendix

\section{The z<1.07 anomaly}
\label{appendixA}
\begin{figure}
    \centering
    \begin{subfigure}[t]{0.5\textwidth}
         \centering
         \includegraphics[width=0.85\textwidth]{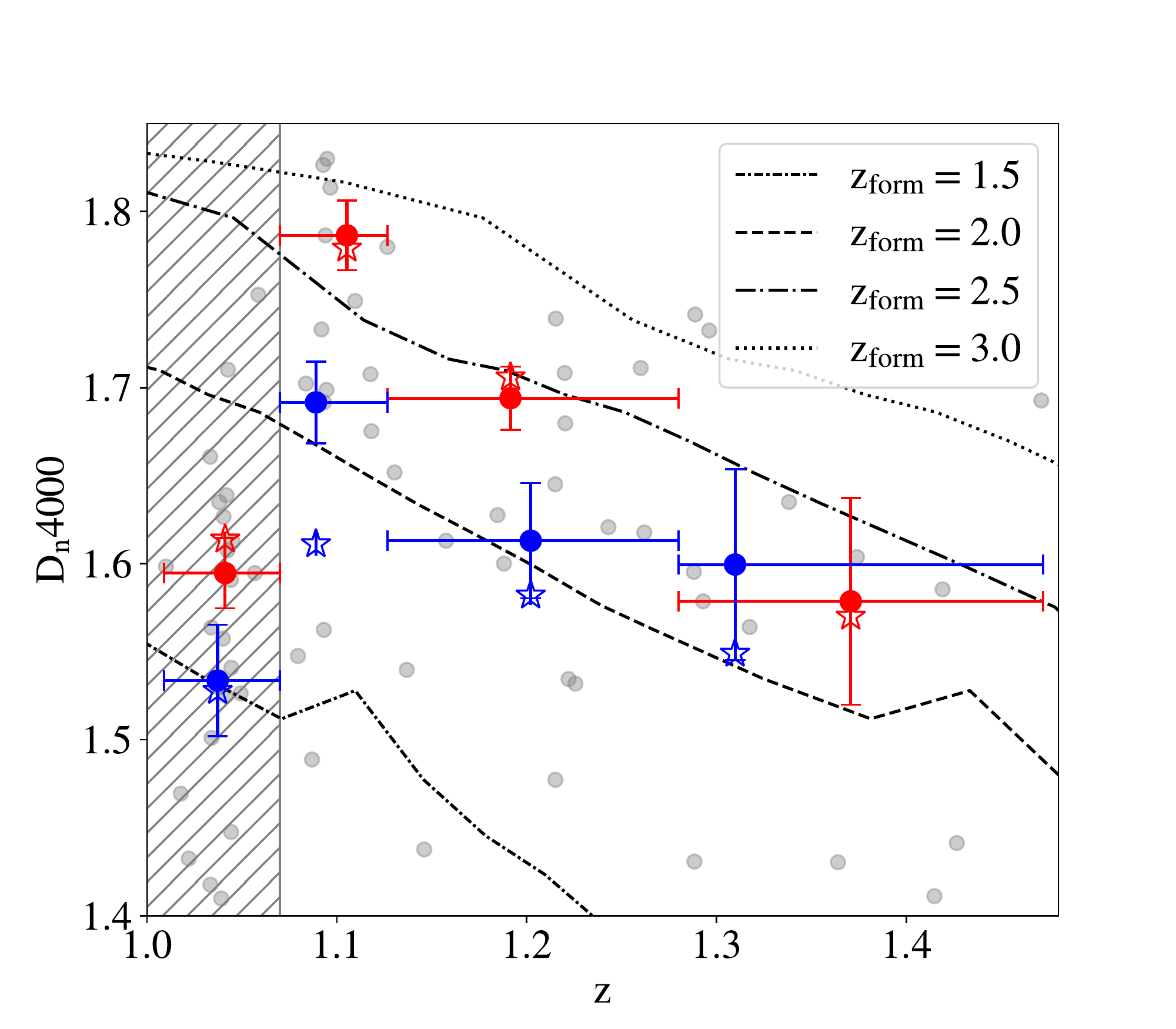}
         \label{fig:Adn4000}
     \end{subfigure}
     \vspace{-0.5cm}
     \caption{$\mathrm{D_n4000}$ trend with redshift. In grey the measurements for the single objects of the CCs sample are shown; blue and red dots are median values for objects with $\mathrm{\mass \leq 10.86}$ (high-mass) and $\mathrm{\mass > 10.86}$ (low-mass) respectively; similarly, blue and red stars are measurements obtained on stacked spectra of low and high-mass samples. In the background, lines refer to evolutionary models from \citet{Bruzual2003} at different redshifts of formation.}\label{fig:App_spec_d4000}
\end{figure}

The discontinuity at 4000 {\AA}, known as D4000 break, is a spectral feature tightly connected to the age (and metallicity) of passive galaxies. For a homogeneous population where the evolution in metallicity is negligible, we expect a decreasing trend with redshift for this quantity \citep{Moresco2012}, as also shown by theoretical models reported in Fig. \ref{fig:App_spec_d4000}.
In the redshift range 1<z<1.07, highlighted in Fig. \ref{fig:App_spec_d4000} with a shaded region, we instead find an anomalous behaviour: all the objects in this redshift bin show a much weaker $\mathrm{D_n4000}$ than expected, comparable to the values obtained at z=1.3-1.4, independently of stellar mass. In order to understand the origin of this anomaly we explored different hypotheses.

\begin{figure}
    \centering
    \includegraphics[width=0.5\textwidth]{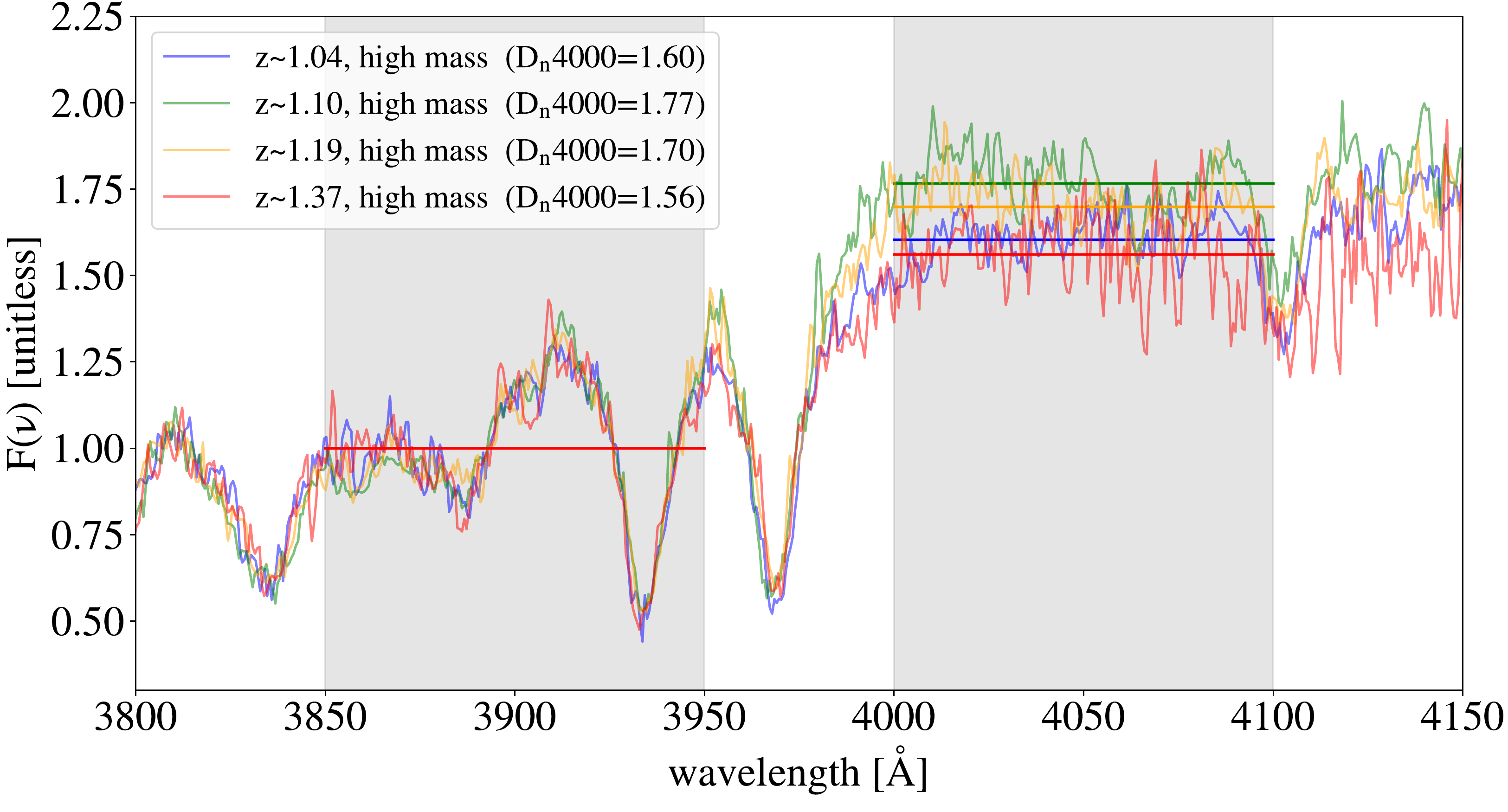}
    \caption{Median stacked spectra obtained for the high-mass sample in each of the four redshift bins, normalised to their mean value in the blue band of the $\mathrm{D_n4000}$. In grey we show the two windows in which the red and blue bands of the $\mathrm{D_n4000}$ are defined, while horizontal lines represent their median values in these two ranges. We remind here that the median spectra have been normalised in the blue window, therefore the average fluxes of the four spectra in that region are identical by definition.}
    \label{fig:App_zoomD4000}
\end{figure}

\smallskip
\textit{Measurement error hypothesis.} The first possible origin that we consider is that some errors in the D4000 measurements were made. To check this, we produce eight stacked spectra, one for each red or blue point in Fig. \ref{fig:App_spec_d4000}, first dividing the sample into a high-mass and a low-mass sample with respect to the median mass ($\mathrm{\massf=10.86}$) and then in four redshift bins. In this way, we can obtain spectra with a higher S/N and, therefore, reduce the possibility of a measurement error. We also normalise the eight stacked spectra to their median value in the range 3850-3950 {\AA}, which is the blue wavelength range used in computing $\mathrm{D_n4000}$. We choose this window because, normalising the spectra in this interval, we are able to visually check in the red band of the $\mathrm{D_n4000}$ eventual differences among the different spectra. The four stacked spectra for the high-mass sample are shown in Fig. \ref{fig:App_zoomD4000}. The coloured horizontal lines at $\mathrm{4000<\lambda<4100 \AA}$ represent the average flux in the red window of the $\mathrm{D_n4000}$, and since the spectra have all been normalised in the blue window, they allow us to check by eye differences in the $\mathrm{D_n4000}$, which is the ratio between the average fluxes in these intervals. As it is possible to see, the average flux at z~1.04 (blue line) is almost equal to the average flux at z~1.34 (red line), while at the other redshifts the behaviour is as expected. We observe the same pattern also for the low-mass stacked spectra. This means that the population of galaxies in the lowest redshift bin, where we find the anomaly, does effectively show a lower $\mathrm{D_n4000}$ than expected, similar to the one obtained in the highest redshift bin.\\
For this reason, we reject the hypothesis of a measurement error.

\smallskip
\textit{Calibration issue hypothesis.} Since this drop in D4000 appears for galaxies at $z<1.07$, we consider the possibility of calibration issues. In fact, it involves all the objects in a specific redshift interval and regards mainly ``wide'' spectroscopic features (like $\mathrm{D_n4000}$ and $\mathrm{Mg{wide}}$, while we don't see this anomaly for spectral indices that concern $\mathrm{\Delta \lambda \lesssim 200}$ {\AA}), so a calibration effect seems plausible. In order to observe this type of behaviour in the D4000, the hypothetical calibration should have increased of $\sim$10\% the observed flux bluer than 8300 {\AA} or, equivalently, decreased it of $\sim$10\% at wavelength redder than 8300 {\AA}, where the D4000 is observed at this redshift. As explained in \citet{Garilli2021}, a correction on the spectra was indeed applied, named \textit{blue-end correction}, but it is of the order of 5\% and significant only at $\mathrm{\lambda \lesssim 5500}$ {\AA} (observed-frame), then not able to justify the observed anomaly. The issue was reported to the VANDELS collaboration, but no other calibration effect was pointed out.\\
Then, a calibration issue can be excluded as the anomaly's source.

\smallskip
\textit{Environmental or selection effects hypothesis.} Up to now we know that the D4000 drop is actually present in the spectra and that it is not related to a measurement issue, so we explore the possibility of it being related to a real effect concerning galaxy properties. A lower D4000 in a passive galaxy could mean a younger age or also a lower metallicity or a lower dust extinction, even if, if this was the case, we should still find a reason for these properties characterising this particular redshift bin. We analyse the available physical properties of the sample, this time splitting it into the two fields of observation, CDFS and UDS, so that also possible anomalies concerning one of the two fields could be pointed out. Looking at the trends in stellar mass, specific star formation rate and dust extinction we don't find substantial differences neither throughout redshift nor among fields. If we consider instead the age as a trigger, we have two possibilities: the first is that this group of galaxies underwent rejuvenation processes, maybe due to an environmental effect; the second is that an unexpected selection effect is present.\\
Looking for an environmental difference in the population at z<1.07, we investigate the presence of overdensities in this redshift interval. Studying the redshift distribution of the sample we observe that both in UDS and CDFS an overdensity is present around z$\sim$1.04. A dense environment can in fact be a reason for rejuvenation if, for example, it triggers intense merger phenomena. Anyway, this should happen equally in both fields to justify the effect in our data, since the D4000 drop is of the same entity in CDFS and UDS. Moreover, we see that the overdensity at z$\sim$1.04 is not the only one, but others are present at higher redshift. At z$\sim$1.09 in UDS, for example, an important overdensity is present but we don't see any anomaly in this redshift bin. For so, the rejuvenation hypothesis seems unlikely.\\
The second option of a possible selection effect was presented to the VANDELS collaboration, but again nothing was found able to justify the anomaly. This hypothesis remains open but, for the moment, there are no arguments supporting it. For this reason, we decide to remove the galaxies at z<1.07 from our sample, to avoid any potential biases in our cosmological analysis.

\section{Assessing the impact of the SFH choice}
\label{appendixB}
The SFH choice is for sure one of the most significant components when building a fitting model since it describes how the star formation rate varies as a function of cosmic time and then dictates the chemical enrichment history of the galaxy, as well as its stellar mass. As discussed in Sect. \ref{sec:3.1FSFwithBagpipes}, the DPL SFH is a valid alternative to the DED SFH that we use as baseline, and actually it has already been adopted while analysing VANDELS data \citep{Carnall2019,Carnall2022}. Two are the main differences with respect to a DED SFH: the first is that a DPL has three free parameters instead of two since its shape is determined separately by the falling ($\alpha$) and the rising slope ($\beta$) of the curve; the second regards the definition of the galaxy age, which is, by default in \texttt{BAGPIPES}, the mass-weighted quantity defined in Eq. \ref{eq:3age_MW} instead of the one in Eq. \ref{eq:3age} adopted for a DED SFH. For the sake of consistency, in this section we use mass-weighted ages for both DED and DPL SFH assumptions. 

\begin{figure*}
    \centering
     \begin{subfigure}[c]{\textwidth}
         \centering
         \includegraphics[width=0.4\textwidth]{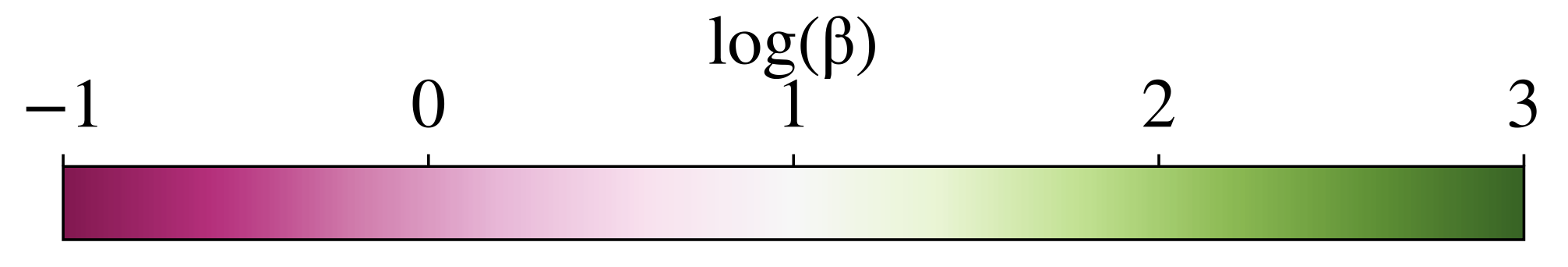}
         \label{fig:legend_dpl}
     \end{subfigure}
     \hfill
     \centering
     \begin{subfigure}[b]{0.45\textwidth}
         \centering
         \includegraphics[width=0.85\textwidth]{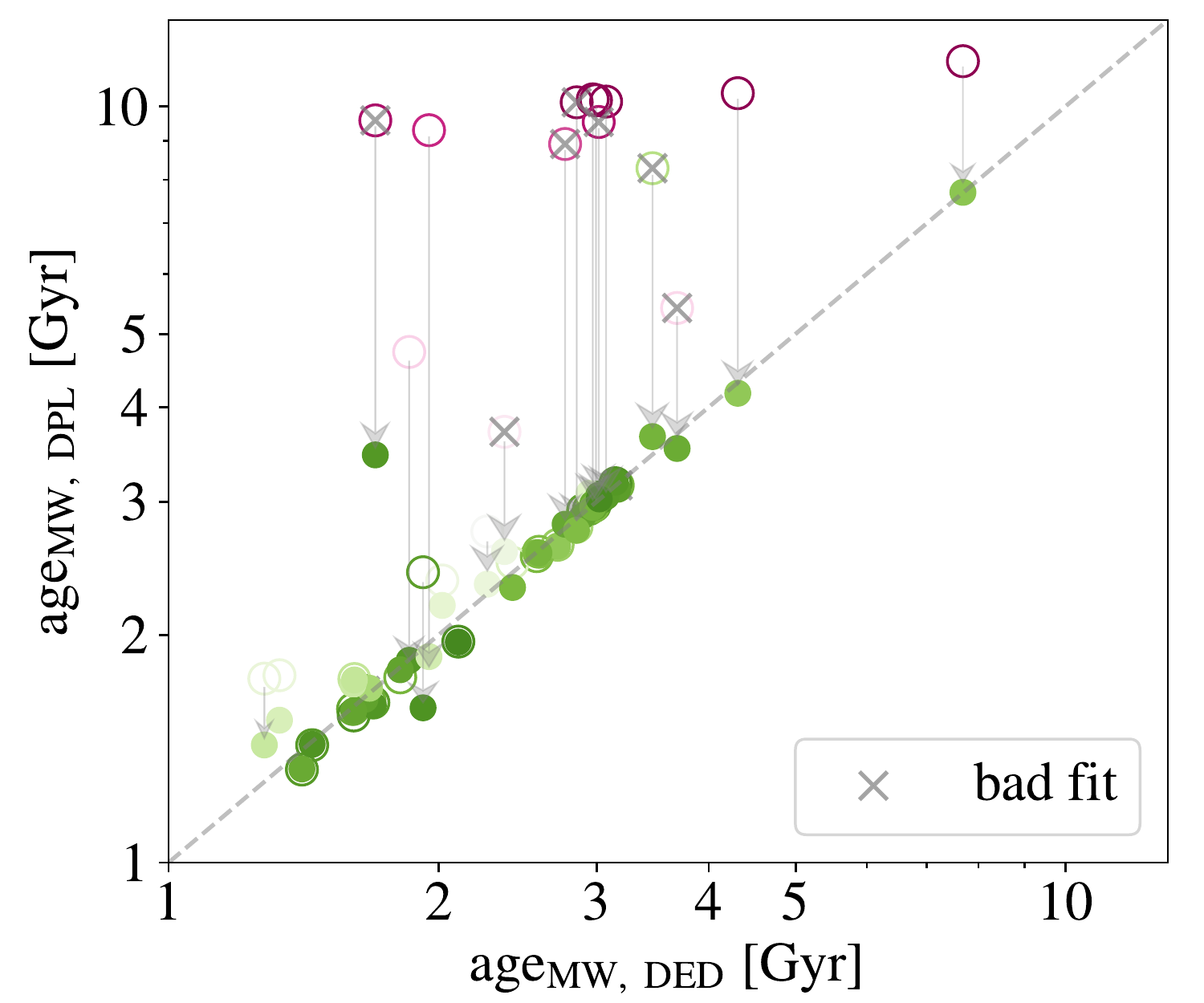}
         \caption{}
         \label{fig:age_age}
     \end{subfigure}
     \hfill
     \begin{subfigure}[b]{0.54\textwidth}
         \centering
         \includegraphics[width=0.7\textwidth]{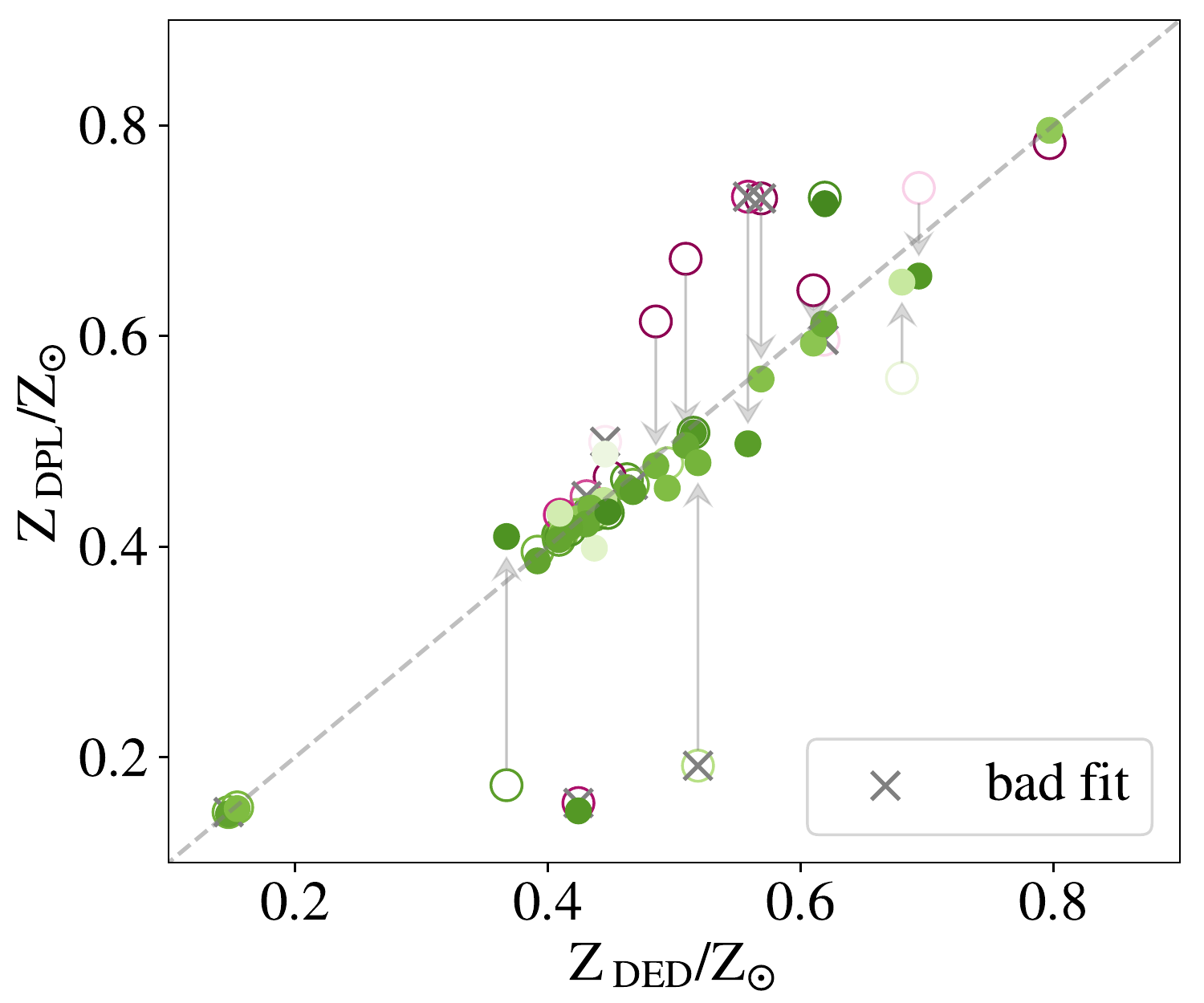}
         \caption{}
         \label{fig:met_met}
     \end{subfigure}
     \hfill
     \begin{subfigure}[b]{0.45\textwidth}
         \centering
         \includegraphics[width=0.85\textwidth]{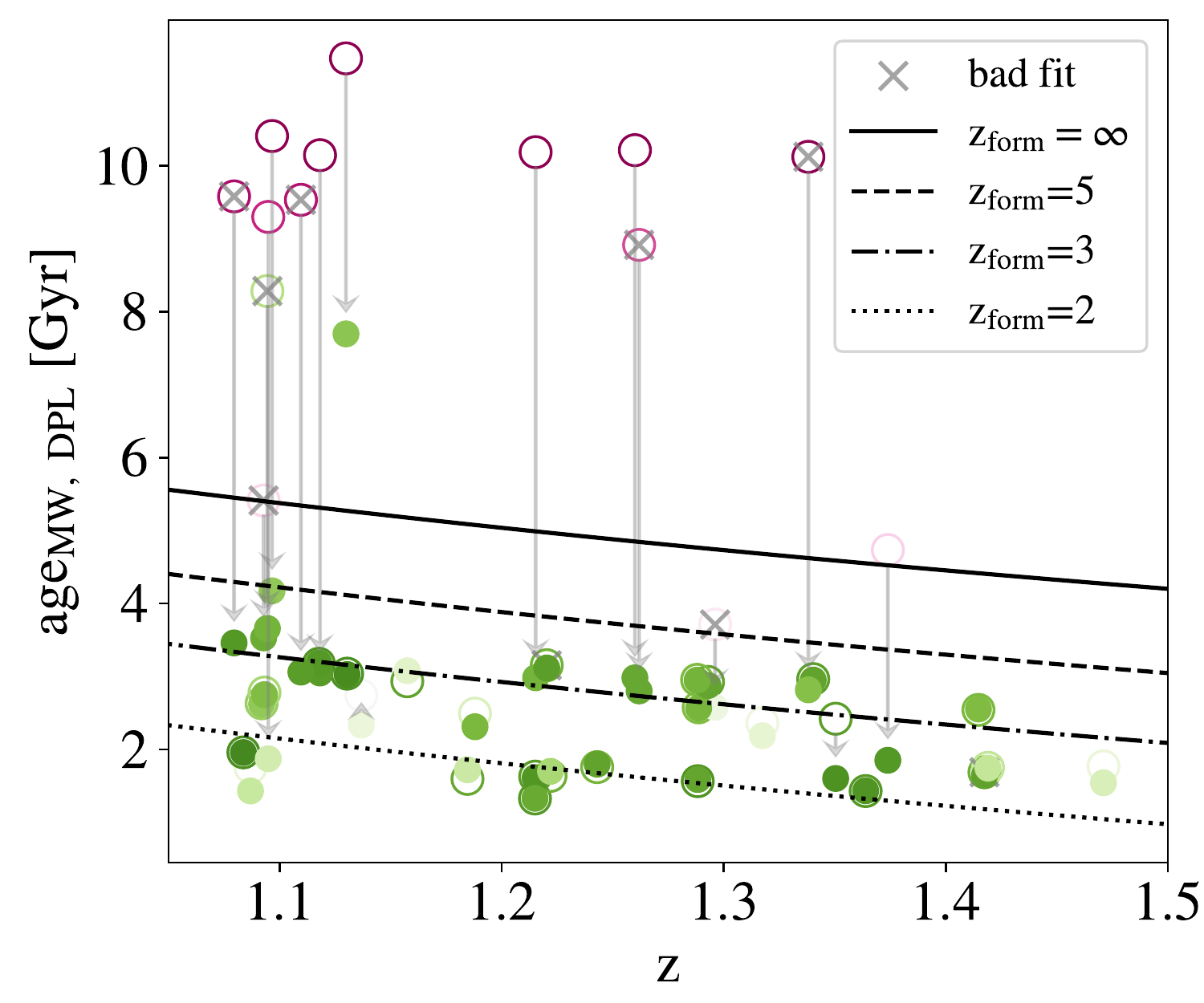}
         \caption{}
         \label{fig:age_z_dpl}
     \end{subfigure}
     \hfill
     \begin{subfigure}[b]{0.54\textwidth}
         \centering
         \includegraphics[width=\textwidth]{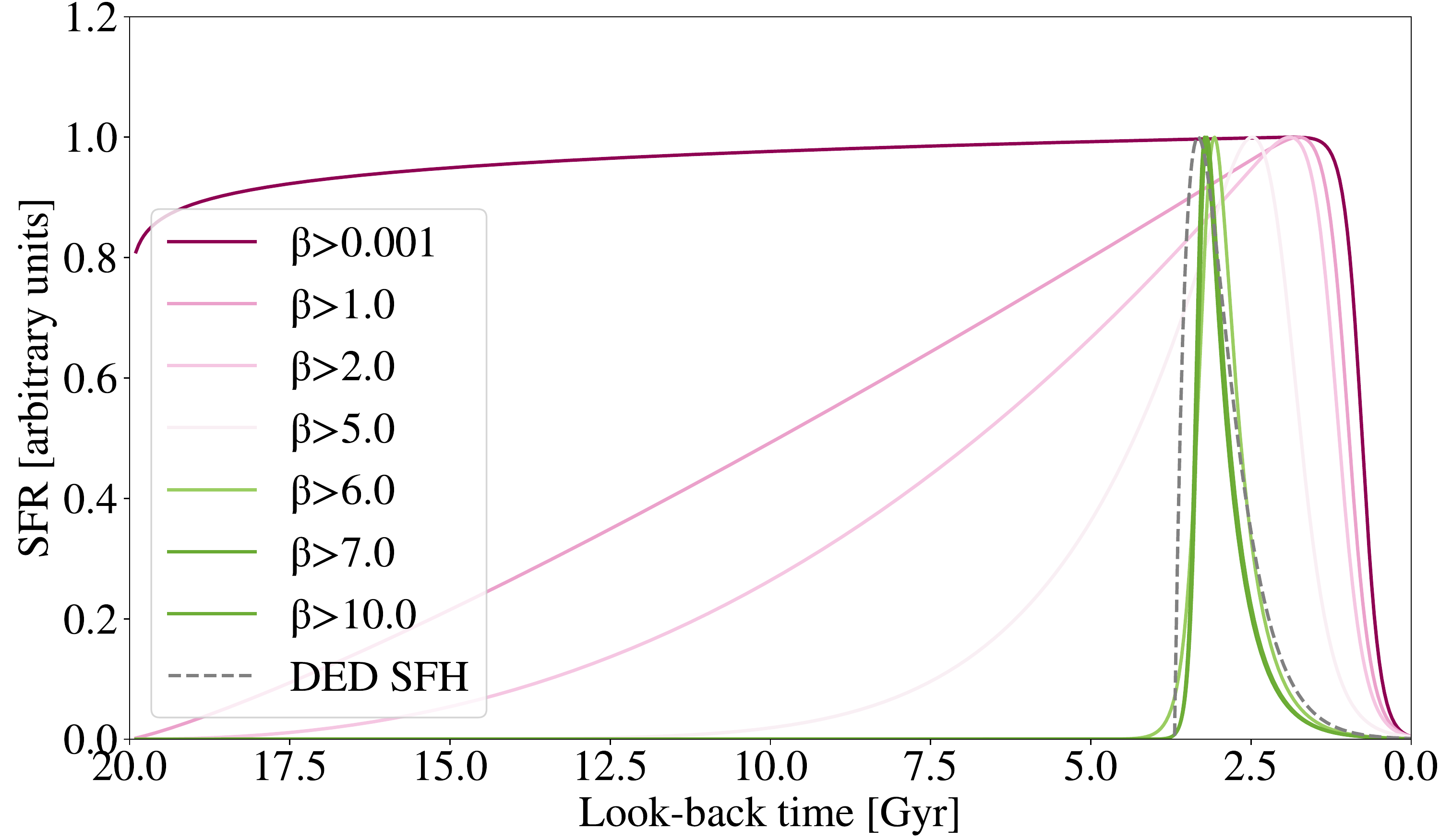}
         \caption{}
         \label{fig:SFH}
     \end{subfigure}
     \hfill
        \caption{Comparison of physical parameters obtained with delayed exponentially declining (DED) and double-power law (DPL) SFH. In panels (a) and (b) we show the comparison of mass-weighted ages and metallicities estimated with the two SFHs. The points are colour-coded according to the parameter $\beta$ of the DPL model, representing the rising slope of the SFH. We notice that the objects with extremely low $\beta$ (represented with open circles) are the ones deviating more from the one-to-one relation. For this reason, we fit again those objects with a more conservative prior on $\beta$ ($\beta>10$), and the grey arrows show how the results converge to the correct relation with the new prior.
        In panel (c) we present the age-redshift relation with the same colour-coding as in the previous panels. In panel (d) the star formation rate is shown as a function of the look-back time normalised to its peak for an object significantly deviating from the one-to-one relation in panel (a). The coloured curves represent how the resulting SFH obtained with the DPL model changes with different priors on $\beta$, colour-coded as presented in the legend; the grey dashed line shows the DED SFH, for comparison. We notice that above $\beta>6$ the DPL model converges almost exactly to the DED model.}
        \label{fig:3dpl_comparison}
\end{figure*}

As presented in Sect. \ref{sec:3.2FSFinVANDELS}, we perform a fit adopting a DPL SFH named configuration 1, with parameters and priors as listed in Tab.~\ref{tab:3priors}. 
Once removed the bad fits, we find an exceptional agreement with baseline results for 75\% of the galaxy population, as shown in Fig. \ref{fig:3dpl_comparison}, with a median percent difference in terms of mass-weighted age, stellar mass, metallicity, dust reddening and velocity dispersion smaller than 1\%. The rest of the sample (six objects) shows a significant difference with respect to baseline results (purple open dots in Fig. \ref{fig:3dpl_comparison}).\\ 
To understand the nature of this difference, we examine all the properties of these six CCs, finding that they show common characteristics, in particular ages larger than 5 Gyr and $\beta$<1. Recalling that $\beta$ sets the rising slope of the SFH, a value $\beta$<1  implies that the first part of the SFH has a concave shape, as in the dark purple curve in Fig. \ref{fig:SFH}, causing an extremely slow rise of the SFR. Solutions of this kind appear highly non-physical because they would require a SFH extremely prolonged over time and without significant episodes of star formation, in contradiction with other indicators available (e.g., the measured sSFR, H/K ratio, etc.) and with all the other results obtained in the literature for these extremely massive and passive systems. For the population whose results agree with baseline ones, instead, the value of $\beta$ spans from a few tens to hundreds, resulting in a very steep rise of the SFR. It seems that relaxing the cosmological prior on ages and, at the same time, adopting a DPL SFH, which is more flexible than a DED SFH in terms of shape, allows the fit, in some cases, to find a combination of parameters leading to a non-physical solution.

We run some tests to understand if a different, physically acceptable solution exists for these objects by increasing the prior on $\beta$. In Fig. \ref{fig:SFH} are shown the SFHs obtained for an example object resulting from seven runs where the $\beta$ prior was gradually raised from 0.001 to 10. We can observe that up to $\beta$>5 the resulting SFH still has an anomalous shape and duration, with values for $\beta$ near the prior limit, suggesting that the fit is still trying to reach for the non-physical solution. From prior $\beta$>6 on, instead, we find solutions with values for $\beta$ two orders of magnitude larger and, above this limit, increasing the value of the prior leads in all cases to the exact same solution, which is also almost identical to the SFH found when fitting with a DED SFH. This means that a physical solution for these objects does exist and that, if the SFH parameters are properly constrained, the resulting SFH has the same shape if either a DED SFH or a DPL SFH is assumed.

Based on these findings, we repeat the fit on all our 49 objects using config. 1 (Tab. \ref{tab:3priors}), this time increasing the lower limit on $\beta$ to 10. In Fig. \ref{fig:age_age} and \ref{fig:met_met} we show the comparison of results for mass-weighted ages and metallicity between baseline (x-axis) and config. 1 (y-axis) both with the old (open dots) and the new prior (full dots). It's clear that after applying the new prior the results of the two fits have a 1-to-1 correlation, with a median percent difference lower than 2\% for both quantities. In terms of stellar mass, velocity dispersion, and dust reddening the discrepancy is even smaller, reaching a median percent difference lower than 1\%. Given the colour code applied in these figures, it is also evident how the results of config. 1 with the old prior are different from the ones obtained with the new prior only when $\beta$ converges to values lower than 1, while in all the other cases they overlap almost perfectly. The same behaviour is noticeable in the age-redshift relation shown in Fig. \ref{fig:age_z_dpl}.

We can conclude that the impact of the SFH choice on the estimation of physical parameters is minimal, under the condition that the SFH parameters are constrained to avoid non-physical solutions. In particular, a lower limit to the rising slope of the DPL SFH $\beta$>10 is effective for this purpose. In addition, the fact that the resulting SFH shape is equivalent when adopting a DED or a DPL SFH further validates the robustness of results obtained with the baseline configuration, despite the existent degeneracy between age and $\tau$ in a DEL SFH discussed in Sect. \ref{sec:3.3PhysProp}.

In addition to these tests, we also used config. 1 results to estimate a value of the Hubble parameter with the DPL SFH, finding very good agreement with the results obtained with the DED SFH. This demonstrates that not only the absolute ages obtained assuming different SFH are in very good agreement, but also the relative ages, proving the robustness of the CC method against the assumption of a specific SFH model (due to the strictness and purity of our selection of CC candidates).

\end{document}